\documentclass[pss]{wiley2sp} 
\usepackage{amsmath}
\usepackage{graphicx}

\begin{document}

\title{Fermionology in Kondo-Heisenberg model: The case of CeCoIn$_{5}$}

\titlerunning{Fermionology in Kondo-Heisenberg model}

\author{%
  Yin Zhong\textsuperscript{\Ast,\textsf{\bfseries 1}},
  Lan Zhang\textsuperscript{\textsf{\bfseries 1}},
  Han-Tao Lu\textsuperscript{\textsf{\bfseries 1}},
  Hong-Gang Luo\textsuperscript{\textsf{\bfseries 1,2}}}
\authorrunning{Yin Zhong et al.}

\mail{e-mail
  \textsf{zhongy05@hotmail.com}, Phone:
  +86-15193133526}

\institute{%
  \textsuperscript{1}\,Center for Interdisciplinary Studies $\&$ Key Laboratory for
Magnetism and Magnetic Materials of the MoE, Lanzhou University, Lanzhou 730000, China\\
  \textsuperscript{2}\,Beijing Computational Science Research Center, Beijing 100084, China\\
  }

\received{XXXX, revised XXXX, accepted XXXX} 
\published{XXXX} 


\abstract{%
%
%
%
\abstcol{%
  Fermi surface of heavy electron systems plays a fundamental role in understanding their variety of puzzling phenomena, for example, quantum criticality, strange metal behavior, unconventional superconductivity and even enigmatic phases with yet unknown order parameters. The spectroscopy measurement of typical heavy fermion superconductor CeCoIn$_{5}$ has demonstrated multi-Fermi surface structure, which has not been in detail studied theoretically in a model system like the Kondo-Heisenberg model. In this work, we make a step toward such an issue with revisiting the Kondo-Heisenberg model. It is surprising to find that the usual self-consistent calculation cannot reproduced the fermionology of the experimental observation of the system due to the unfounded sign binding between the hopping of the conduction electrons and the mean-field valence-bond order.
  }{%
   To overcome such inconsistency, we assume that the sign binding should be relaxed and the mean-field valence-bond order can be considered as a free/fit parameter so as to meet with real-life experiments. Given the fermionology, the calculated effective mass enhancement, entropy, superfluid density and Knight shift are all in qualitative agreement with the experimental results of CeCoIn$_{5}$, which confirms our assumption. Our result supports a $d_{x^{2}-y^{2}}$-wave pairing structure in heavy fermion material CeCoIn$_{5}$. In addition, we have also provided the scanning tunneling microscopy (STM) spectra of the system, which is able to be tested by the present STM experiments.}}

%
%

\maketitle   
\section{Introduction} \label{intr}
Elucidating the structure of Fermi surface is a key step to understanding the nature of strongly correlated electron systems. Experimentally, the Fermi surface can be measured by powerful spectroscopy techniques, e.g. angle-resolved photoemission spectroscopy (ARPES) and quasiparticle interference (QPI) of scanning tunneling microscopy (STM),\cite{Shen2003,Renner2007,Hoffman2011} which provide large number of direct information on the fermionology of high temperature superconducting cuprate and pnitidate.\cite{Lee2006,Armitage2010,Stewart2011,Scalapino2012,Chubukov2012} Recently, these state-of-art techniques has been successfully used to probe local/momentum space electronic structure of several heavy fermion compounds, such as URu$_{2}$Si$_{2}$, YbRh$_{2}$Si$_{2}$ and CeCoIn$_{5}$.\cite{Koitzsch2009,Jia2011,Davis2010,Wirth2011,Yazdani2012,Davis2013,Yazdani2013,Morr2014} Particularly, for the quasi-two-dimensional heavy fermion superconductor CeCoIn$_{5}$,\cite{Monthoux2001,Pfleiderer2011} both ARPES and QPI experiments reveal a hole-like Fermi pocket around $(0,0)$ and one or more electron-like ones centered at $(\pi,\pi)$ above the superconducting critical temperature $T_{c}\sim 2.3$K. \cite{Koitzsch2009,Jia2011,Yazdani2012,Morr2014}

However, many previous works \cite{Watanabe2007,Assaad2008,Zhang2011,Vekhter2013,Hoshino2013,Asadzadeh2013,Liu2014} have focused on the case with a single large Fermi surface around $(\pi,\pi)$, which is obviously inconsistent with the fermionology of the experimentally observation of the systems, like CeCoIn$_{5}$. This motivates us to check the physics involved multi-Fermi surface. Theoretically, the Kondo lattice or Kondo-Heisenberg model \cite{Senthil2003,Senthil2004} is believed to be able to capture generic Fermi surface structure of such system, which originates from Kondo hybridization between conduction electron sea and local spins.\cite{Doniach,Hewson,Tsunetsugu,Rosch,Coleman2010} In addition, the short-ranged magnetic interaction, which results from the well-known Ruderman-Kittel-Kasuya-Yosida (RKKY) exchange interaction, is also able to lead to more complicated topology of Fermi surface and even to induce radical change of Fermi surface, namely the Lifshitz transition.\cite{Zhang2011,Vekhter2013,Hoshino2013,Liu2014}

With the well-established large-N mean-field theory, we find that the self-consistent solution of mean-field equations is unlikely to give rise to the desirable multi-Fermi surface structure due to an unexpected and unfound previously sign binding, namely, the sign of valence bond order (kinetic energy) of local spins is locked into the sign of conduction electron hopping $t$. Surprisingly, such sign binding is also true in other many-body models  and we guess that such sign binding is an universal feature in the fermionic large-N mean-field theory, although a general proof is absent but for a simple quantum XY model the analytical result is provided (see Appendix). This feature has not been noticed in previous studies and is a new finding of the present paper. Therefore, we have to relax the self-consistency of valence bond order, which actually leads to well-defined electron and hole Fermi surface without elaborate tuning. We should emphasize that valence bond order is also treated as an external free parameter or fitting parameter in the recent experimental analysis on momentum space structure of normal and superconducting states for CeCoIn$_{5}$.\cite{Davis2013,Morr2014} Importantly, the calculated Fermi surface is qualitative similar to the findings in spectroscopy experiments, thus confirms the validity of our theoretical model calculation and physical arguments. Interestingly, the calculated effective mass is well consistent with the quantum oscillation measurement in CeCoIn$_{5}$. Furthermore, for the given fermionology, we calculate the entropy, superfluid density and Knight shift in the possible unconventional superconducting state, which are in qualitative agreement with those experimental observations in heavy fermion superconductor CeCoIn$_{5}$.
The well agreement with those experiments verifies the rationality of our core assumption that the valence-bond order can be considered as a free parameter, which gives
the observed dispersion of local electrons.\cite{Davis2013}
In addition, we also provide the STM spectra of this two-band system, which shows crucial quantum interference effect between conduction and local electron paths. The lineshape of STM differential conductance shows characteristic Fano resonance when tunneling is dominated by conduction electrons while a large zero energy peak appears if local electrons are more active. We hope that the present work may be helpful for further understanding on the complicated Fermi surface topology of heavy electron systems and the corresponding anomalous behaviors.

The remainder of this paper is organized as follows. In Sec. \ref{sec1}, we first introduce the Kondo-Heisenberg model on the square lattice and the corresponding mean-field equations are derived. Then Sec. \ref{sec2} is devoted to the mean-field solution. It is found that we have to treat valence bond order as a free parameter so as to realize the multi-Fermi surface structure. In Sec. \ref{sec3}, some physical quantities like energy gap, entropy, and superfluid density are calculated in the superconducting state and compared qualitatively with the experimental observations. In Sec. \ref{sec4}, the STM spectra are presented, and we focus on the quantum interference effect between conduction and local electrons. In Sec. \ref{sec5}, we make some interesting extensions on non-Fermi liquid normal state and the issue on superfluid density. Finally, Sec. \ref{sec6} is devoted to a brief conclusion.

\section{Kondo-Heisenberg model}\label{sec1}
The Kondo-Heisenberg model we considered is standard, which reads,\cite{Zhang2011,Senthil2003}
\begin{eqnarray}
H=\sum_{k\sigma}\varepsilon_{k}c_{k\sigma}^{\dag}c_{k\sigma}+J_{K}\sum_{i}S_{i}^{c}\cdot S_{i}^{f}+J_{H}\sum_{<i,j>}S_{i}^{f}\cdot S_{j}^{f} \nonumber
\end{eqnarray}
where the conduction electron has energy spectrum $\varepsilon_{k}=-2t(\cos k_{x}+\cos k_{y})+4t'\cos k_{x}\cos k_{y}-\mu$ with chemical potential $\mu$. The local spins are denoted by fermionic representation $S_{i}^{f}=\frac{1}{2}\sum_{\sigma\sigma'}f_{i\sigma}^{\dag}\tau_{\sigma\sigma'}f_{i\sigma'}$
with $\tau$ being the standard Pauli matrices. In addition, the local constraint $\sum_{\sigma}f_{i\sigma}^{\dag}f_{i\sigma}=1$ should be fulfilled at each site to prohibit any charge fluctuation. Physically, this Hamiltonian describes two competing tendencies: One is the Kondo screening, which leads to the formation of collective spin-singlet state among local moments and conduction electrons. The other is the short-ranged antiferromagnetic fluctuation reflected by the explicitly introduced Heisenberg interaction between local moments. The complicated phenomena in diverse heavy electron systems is believed to be captured by these two active factors.
Usually, to get the qualitatively correct information in the paramagnetic heavy fermion liquid state, the fermionic large-N or slave-boson mean-field theory is
widely utilized.\cite{Read1983,Coleman1987} Here, we will follow the treatment of the fermionic large-N mean-field theory to get an effective Hamiltonian.
However, we should emphasize that similar effective Hamiltonian can also be obtained from more phenomenological Fermi liquid assumption. Therefore, in some sense,
when facing to realistic experimental data, the effective Hamiltonian is more useful than the original microscopic model and any parameter in the effective Hamiltonian
should be treated as effective parameters rather than mean-field parameters or microscopic parameters. We will hold this point if comparing with experiments is involved in the remaining parts of this paper.

After performing the standard large-N mean-field approximation,\cite{Lee2006,Zhang2011} the resultant Hamiltonian reads
\begin{eqnarray}
H&&=\sum_{k\sigma}[\varepsilon_{k}c_{k\sigma}^{\dag}c_{k\sigma}+\chi_{k}f_{k\sigma}^{\dag}f_{k\sigma}+\frac{J_{K}V}{2}(c_{k\sigma}^{\dag}f_{k\sigma}+f_{k\sigma}^{\dag}c_{k\sigma})]\nonumber\\
&&+E_{0}.\label{eq1}
\end{eqnarray}
Here, local spins acquire dissipation $\chi_{k}=J_{H}\chi \eta_k + \lambda$ with $\eta_k = \cos k_{x}+\cos k_{y}$ due to the formation of valence-bond order $\chi=\sum_{\sigma}\langle f_{i\sigma}^{\dag}f_{j\sigma}\rangle$. Physically, such valence-bond order reflects the quantum dynamics of local spins, which competes with magnetic long-ranged order. Lagrangian multiplier $\lambda$ is introduced to impose the local constraint on average. Meanwhile, Kondo screening effect is encoded by the hybridization between conduction electron and local spins via $V=-\sum_{\sigma}\langle c_{i\sigma}^{\dag}f_{j\sigma}\rangle$. In addition, there is a constant energy shift $E_{0}=N_{s}[J_{K}V^{2}/2+J_{H}\chi^{2}-\lambda+\mu n_{c}]$ with the number of lattice sites $N_s$ and the occupied number of $n_c$ of conduction electrons. This constant energy should be added when free energy or ground-state energy is considered.

In previous studies,\cite{Zhang2011,Vekhter2013,Hoshino2013,Liu2014,Watanabe2007,Assaad2008,Asadzadeh2013} a large Fermi surface around $(\pi,\pi)$ is discovered and the increasing of Heisenberg interaction $J_{H}$ leads to appearance of several small Fermi pockets. However, no hole-like Fermi surface emerges
around $(0,0)$, thus it is unable to contact with the spectroscopy experiments on quasi-two-dimensional heavy fermion superconductor CeCoIn$_{5}$.\cite{Koitzsch2009,Jia2011,Yazdani2012,Morr2014} But, it is easy to see that if the hopping parameter $t$ of conduction electron is negative, the desirable hole-like Fermi surface can be obtained. In contrast, if a positive $t$ is used, as done in previous works, \cite{Zhang2011,Vekhter2013,Hoshino2013,Liu2014,Watanabe2007,Assaad2008,Asadzadeh2013} one can obtain a large Fermi surface around $(\pi,\pi)$.
Therefore, throughout the present paper, we only consider the case of $t<0$.

\subsection{Mean-field equations}
The mean-field Hamiltonian of Eq. (\ref{eq1}) can be diagonalized by the following transformation,
\begin{eqnarray}
&&c_{k\sigma}=\alpha_{k}A_{k\sigma}-\beta_{k}B_{k\sigma}\nonumber \\
&&f_{k\sigma}= \beta_{k}A_{k\sigma}+\alpha_{k}B_{k\sigma}\label{eq2}
\end{eqnarray}
with $\alpha_{k}^{2}=\frac{1}{2}(1+\frac{\varepsilon_{k}-\chi_{k}}{E_{0k}})$, $\beta_{k}^{2}=\frac{1}{2}(1-\frac{\varepsilon_{k}-\chi_{k}}{E_{0k}})$ and $\alpha_{k}\beta_{k}=\frac{J_{K}V}{2E_{0k}}$.
Here, we have defined $E_{0k}=\sqrt{(\varepsilon_{k}-\chi_{k})^{2}+(J_{K}V)^{2}}$.

Then, the original Hamiltonian Eq. (\ref{eq1}) reads
\begin{eqnarray}
H=\sum_{k\sigma}[E_{k}^{+}A_{k\sigma}^{\dag}A_{k\sigma}+E_{k}^{-}B_{k\sigma}^{\dag}B_{k\sigma}]+E_{0},\label{eq3}
\end{eqnarray}
where the quasiparticle energy $E_{k}^{\pm}=\frac{1}{2}(\varepsilon_{k}+\chi_{k}\pm E_{0k})$. So, the corresponding free energy is
\begin{eqnarray}
F=-2T\sum_{k}[\ln(1+e^{E_{k}^{+}/T})+\ln(1+e^{E_{k}^{-}/T})]+E_{0}\label{eq4}
\end{eqnarray}
and four self-consistent equations are derived from the condition $\frac{\partial F}{\partial V}=\frac{\partial F}{\partial \chi}=\frac{\partial F}{\partial \lambda}=\frac{\partial F}{\partial \mu}=0$.
\begin{eqnarray}
&&J_{K}\sum_{k}\frac{f_{F}(E_{k}^{+})-f_{F}(E_{k}^{-})}{E_{0k}}=-1\nonumber \\
&&\sum_{k}\eta_k (\alpha_{k}^{2}f_{F}(E_{k}^{-})+\beta_{k}^{2}f_{F}(E_{k}^{+}))=-\chi\nonumber \\
&&2\sum_{k}(\alpha_{k}^{2}f_{F}(E_{k}^{-})+\beta_{k}^{2}f_{F}(E_{k}^{+}))=1\nonumber \\
&&2\sum_{k}(\alpha_{k}^{2}f_{F}(E_{k}^{+})+\beta_{k}^{2}f_{F}(E_{k}^{-}))=n_{c},\label{eq5}
\end{eqnarray}
where $f_F(x)$ denotes the Fermi distribution function.
\subsection{Self-consistent solution and sign binding}
\begin{figure}[tbp]
\includegraphics[width = 0.45\columnwidth]{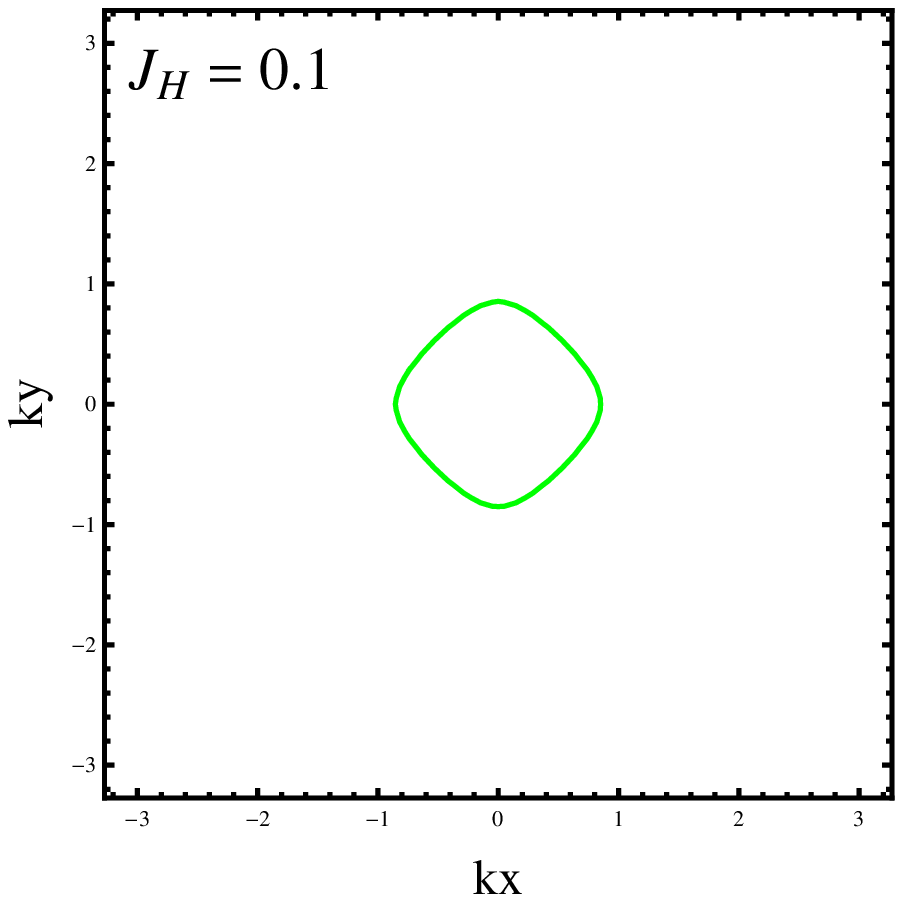}
\includegraphics[width = 0.45\columnwidth]{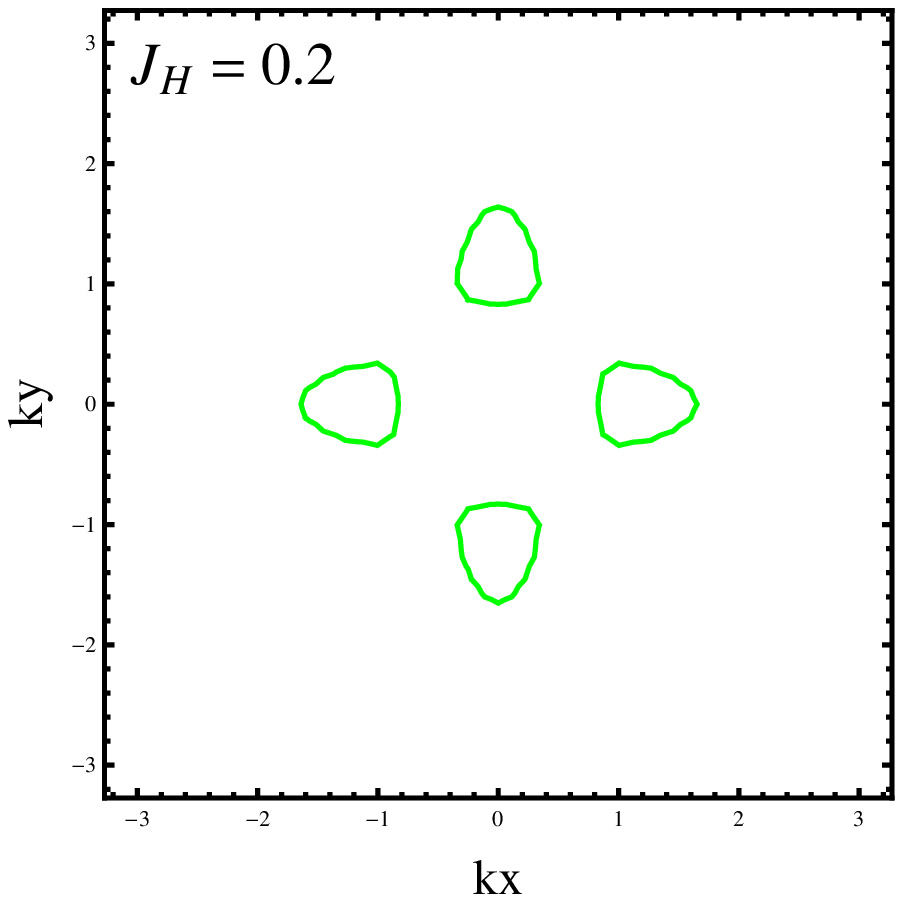}
\includegraphics[width = 0.45\columnwidth]{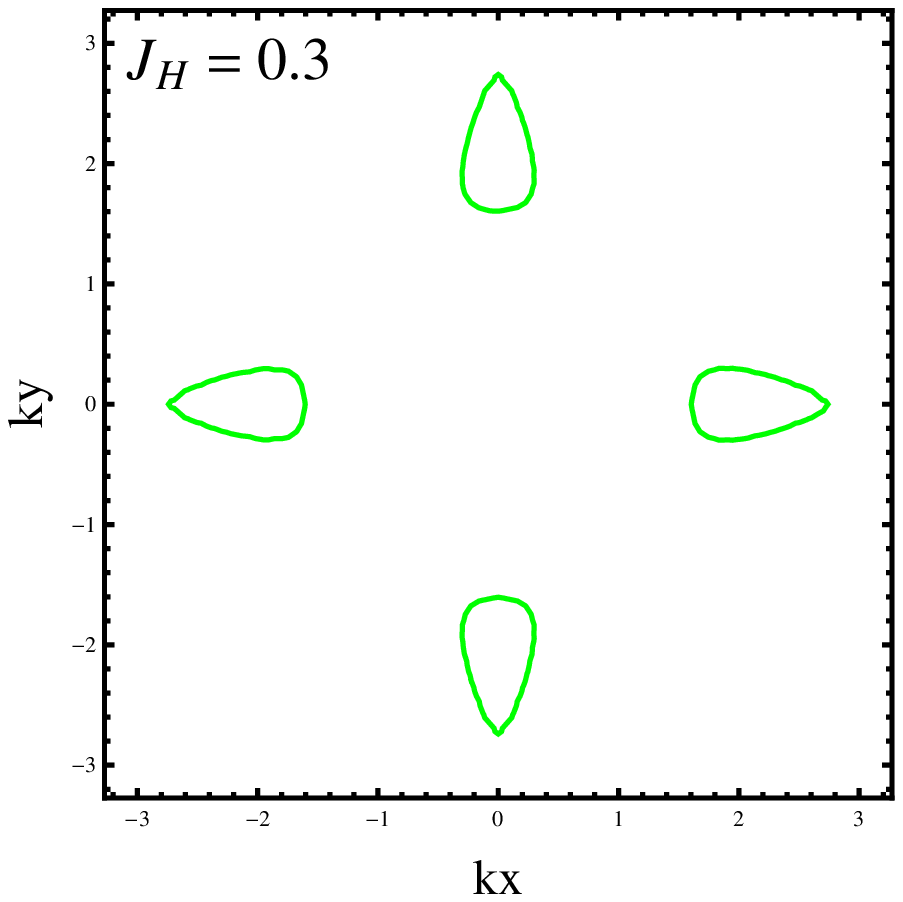}
\includegraphics[width = 0.45\columnwidth]{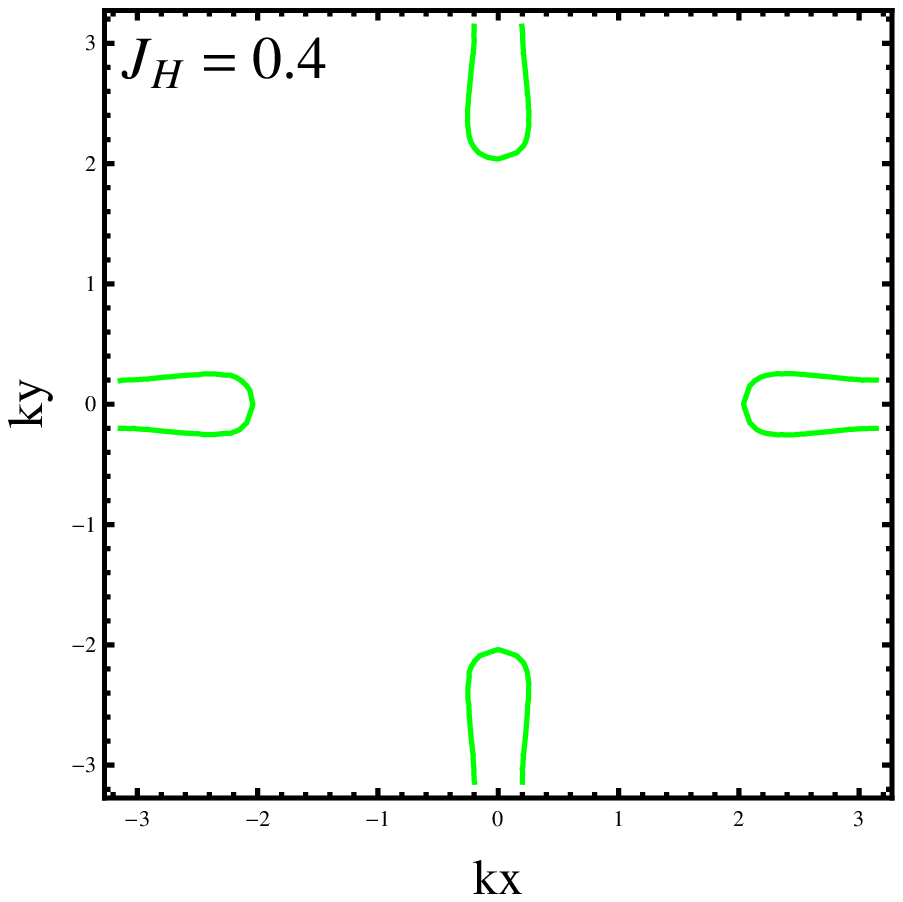}
\caption{\label{fig1} Evolution of Fermi surface with increasing $J_{H}=0.1,0.2,0.3,0.4$. Only hole-like Fermi surface pockets have been obtained in self-consistent calculations.}
\end{figure}

With self-consistent equations in hand, it is ready to obtain some useful physical quantities like the structure of Fermi surface.
Here, we plot evolution of Fermi surface with increasing $J_{H}$ in Fig.\ref{fig1}. Without loss of generality, the parameters are setting to $t=-1,t'=-0.3,J_{K}=2,n_{c}=0.9,T=0.001$ and $J_{H}=0.1,0.2,0.3,0.4,0.5,0.6$. Thus, the studied system is well inside the Kondo-screened regime where magnetic fluctuations and the resulting corrections to the mean-field level are expected to be weak.\cite{Read1983,Paul2007}
Other cases with different doping level like $n_{c}=0.8,0.85,0.95$ are checked and no qualitative changes appear.
Apparently, from Fig. \ref{fig1}, one can see that generally there is no hole-like Fermi surface emerges
around $(0,0)$ if the short-ranged antiferromagnetic (Heisenberg) interaction is involved. In other words, the Heisenberg interaction breaks the hole-like Fermi pocket at $(0,0)$ into four small hole Fermi surface around $(\pi,0)$ and its equivalent points. Furthermore, no electron-like Fermi surface emerges near $(\pi,0)$, which is contrast to the existing experimental data of CeCoIn$_{5}$.

\begin{figure}[tbp]
\centering
\includegraphics[width = 0.45\columnwidth]{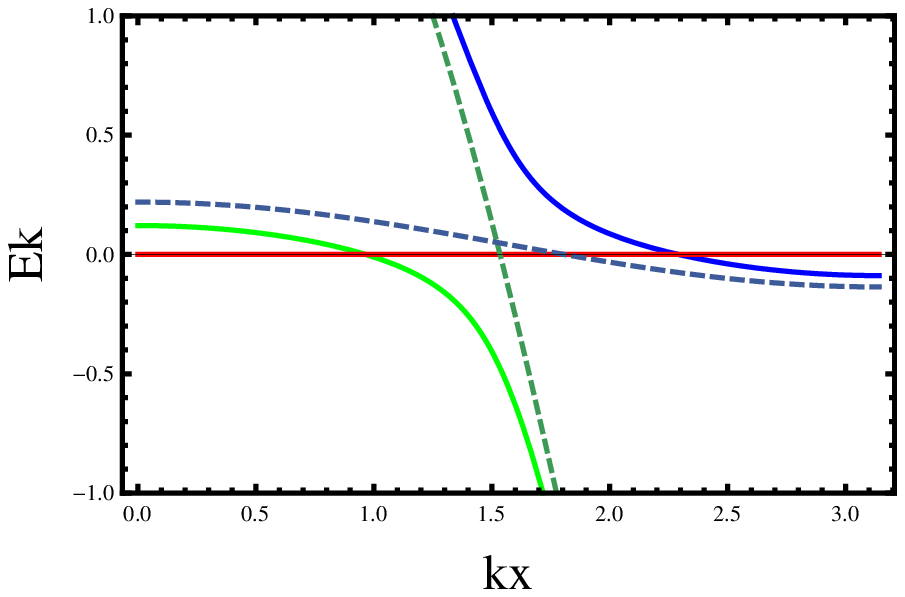}
\includegraphics[width = 0.45\columnwidth]{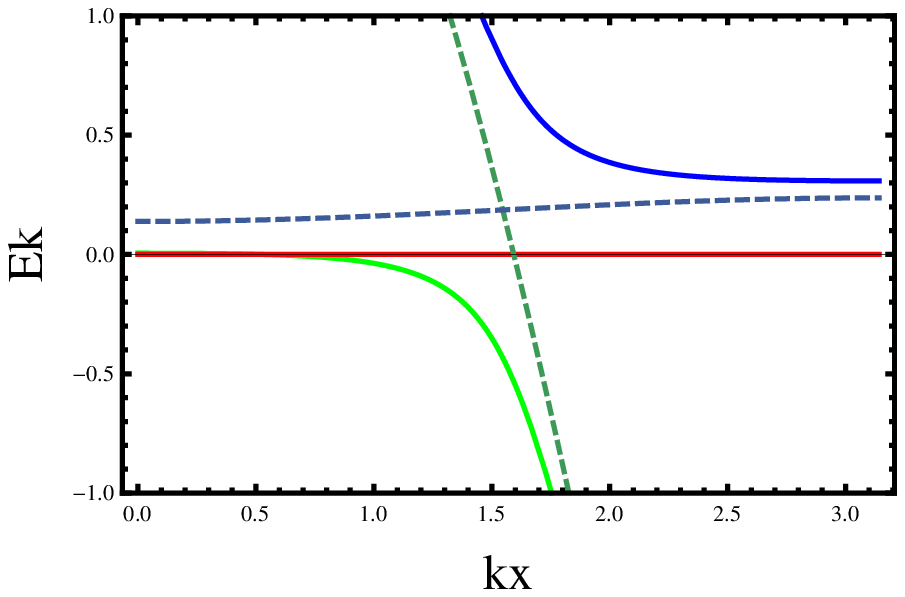}
\caption{\label{fig2}  The hybridization quasiparticle band (blue and green solid lines) for $\chi>0$ (left) versus $\chi<0$ (right) along the direction $(0,0)$ to $(\pi,\pi)$. The dashed lines represent the bare bands of the local and conduction electrons. For $\chi>0$ it is possible to obtain both hole- and electron-like fermi pockets, which is in contrast to that of $\chi < 0$.}
\end{figure}

Actually, these unpleasant results are caused by the mismatch between the sign of hopping $t$ and the valence bond order $\chi$ as what can be seen in Fig. \ref{fig2}. In Fig. \ref{fig2}, the left one shows the quasiparticle band with $\chi>0$ and a hole and electron-like Fermi surface are clearly observed. In contrast, if $\chi<0$, which has the same sigh as $t$ ,it is not possible to have both a hole and electron-like Fermi surface. It should be emphasized that the former case is obtained if we treat $\chi$ as a free parameter, which may be determined by weak hopping of local electrons, rather than a self-consistent mean-field parameter as in the latter case.
More importantly, we have searched a large number of parameters to find solutions with a positive valence bond order but it turns out that the sign binding $sgn(t)=sgn(\chi)$ is always true if $\chi$ is self-consistently determined by mean-field equations. We should emphasize that this feature is not noticed in previous studies and is a new finding of the present paper. Therefore, we have to conclude that if we want to reproduce the observed multi-Fermi surface structure, the self-consistency of valence bond order $\chi$ has to be relax, which means $\chi$ is just an external free parameter like $t$ and $J_{H}$.

As a matter of fact, the valence bond order $\chi$, which gives the local electron dispersion, is also treated as an external free parameter or fitting parameter in recent experimental analysis on momentum space structure of normal and superconducting states for CeCoIn$_{5}$.\cite{Davis2013,Morr2014} In their work, the Hamiltonian Eq.\ref{eq1} is from a conceived extended Kondo-Heisenberg model, where the Heisenberg-like interaction results from the fitting to the observed quasiparticle energy band and quasiparticle interference pattern. In their perspective, the dispersion of local electrons should result from the non-local RKKY exchange interaction and is
an essential element to understand the observed renormalized quasiparticle band. Above all, they think that the origin of the dispersion of local electrons is intrinsic at least in materials like CeCoIn$_{5}$ and should not be considered as an induced effect from true order like valence bond order.\cite{Davis2013,Morr2014}
Therefore, we may consider $\chi$ as a free/fit parameter, which reflects the weak but important dispersion of local electrons. Without this dispersion, the multi-Fermi surface is unable to be obtained as what can be seen in the next section.

\section{Fermi surface with positive valence bond order} \label{sec2}
\begin{figure}[tbp]
\centering
\includegraphics[width = 0.45\columnwidth]{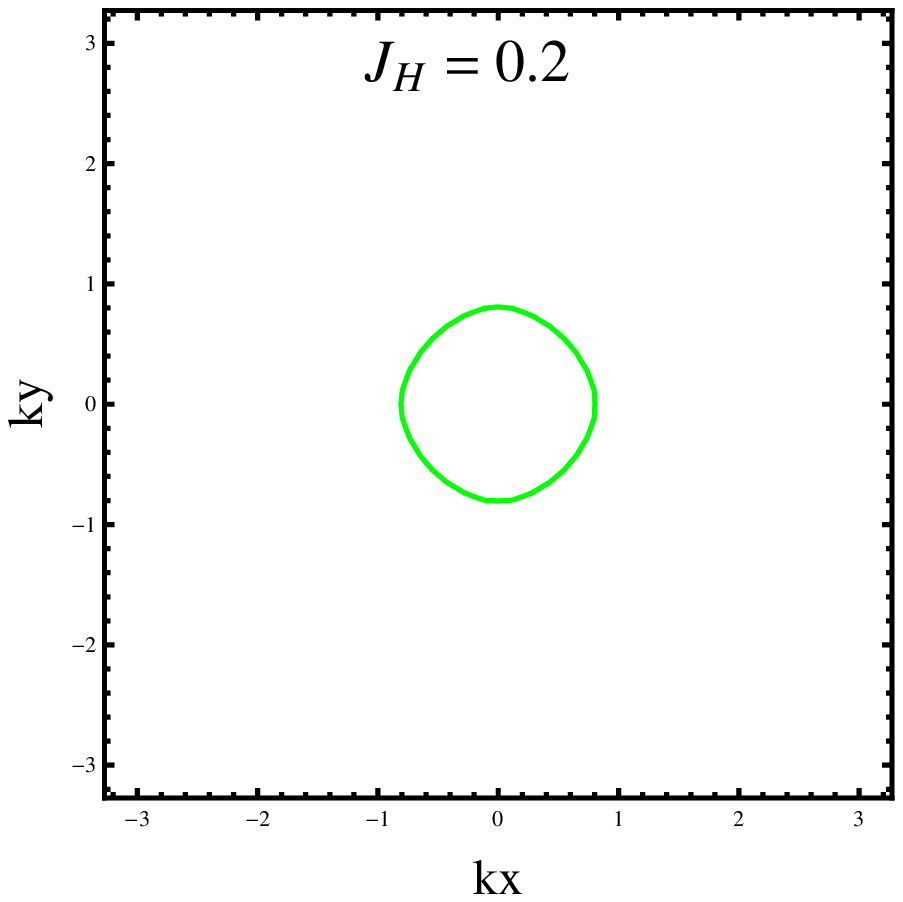}
\includegraphics[width = 0.45\columnwidth]{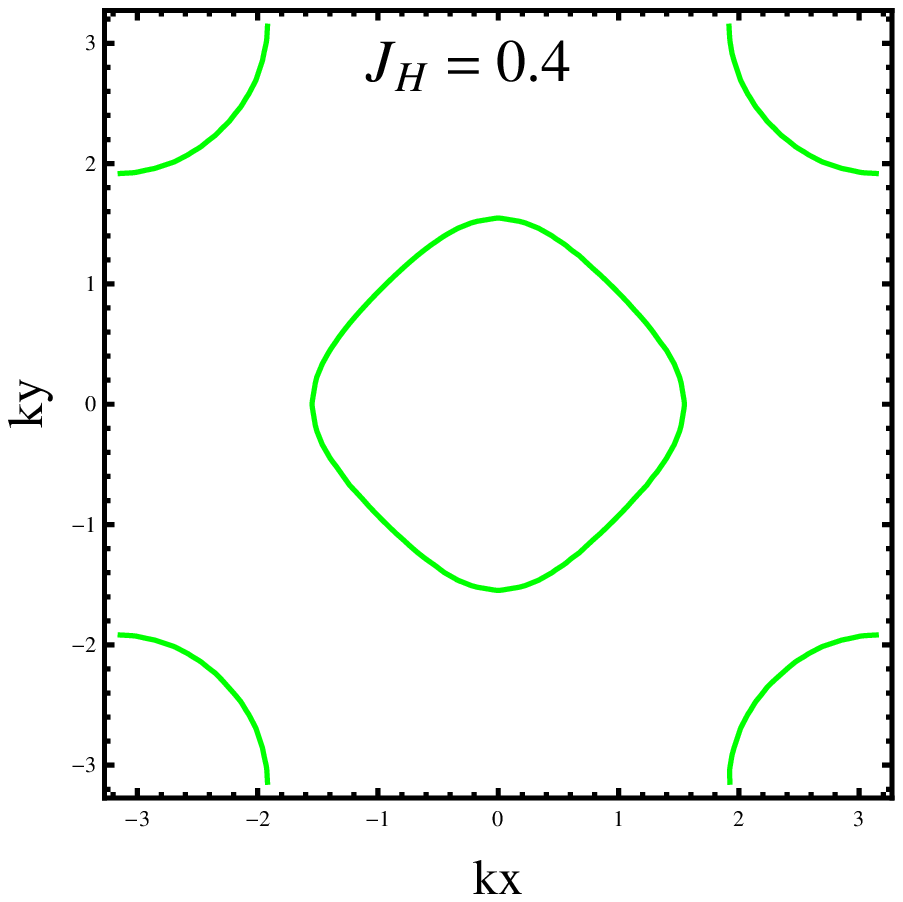}
\includegraphics[width = 0.45\columnwidth]{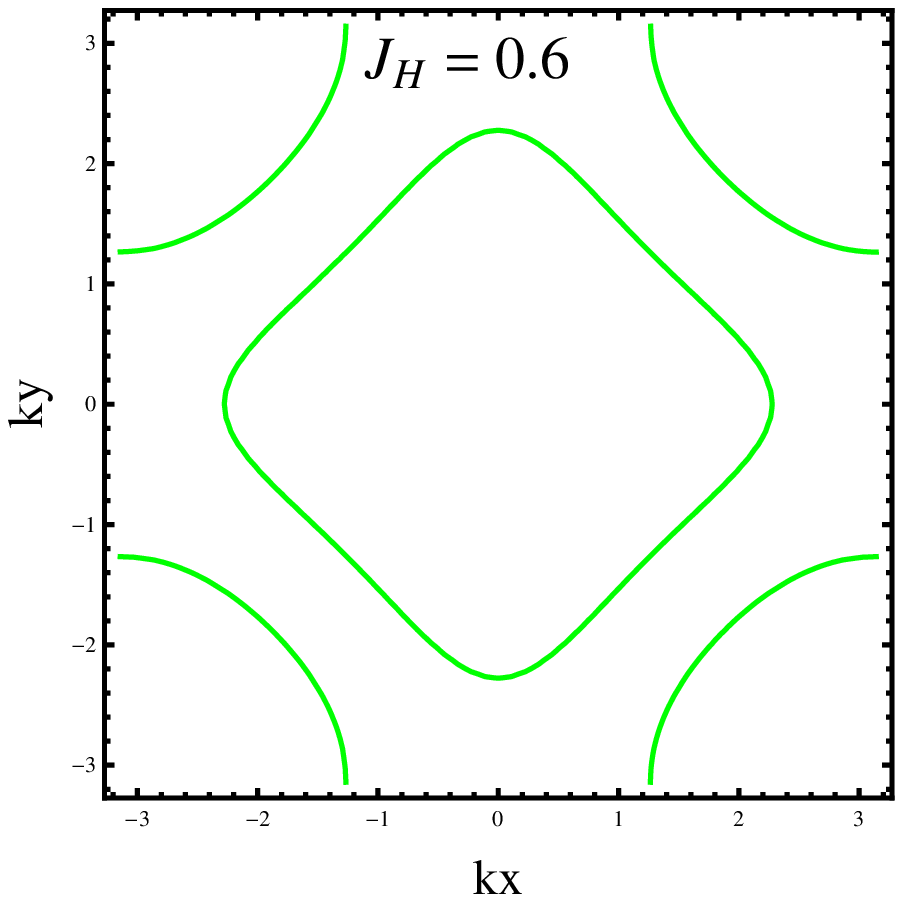}
\includegraphics[width = 0.45\columnwidth]{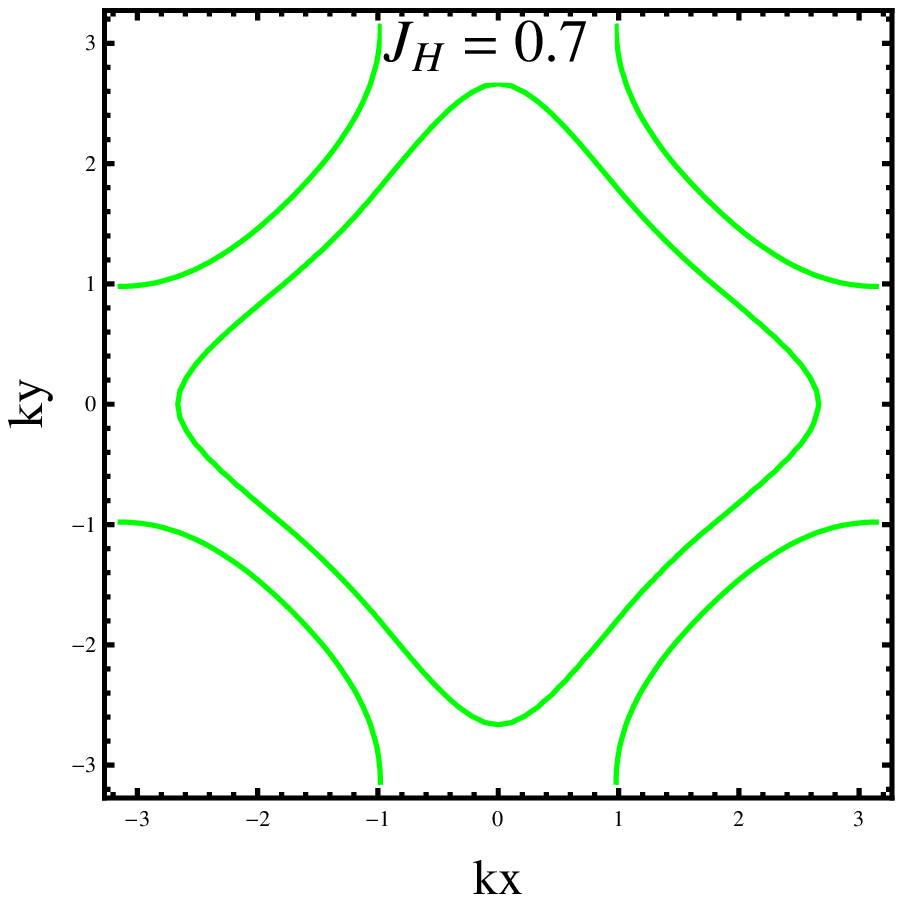}
\caption{\label{fig3} The evolution of Fermi surface with increasing $J_{H}=0.2,0.4,0.6,0.7$.}
\end{figure}

As discussed in last section, due to the unwanted sign structure, we have to relax the self-consistency of $\chi$, thus only three mean-field equations are needed in stead of four. Then, we plot the structure of Fermi surface with typical parameters $t=-1,t'=-0.3,J_{K}=2,n_{c}=0.9,T=0.001,\chi=0.2222$ and $J_{H}=0.2,0.4,0.6,0.7$ in Fig.\ref{fig3}. [We use $\chi=0.2222$ here since in the mean-field solution, this value is generically valid versus different doping level and Heisenberg interaction.  ]

\begin{figure}[tbp]
\centering
\includegraphics[width = 0.45\columnwidth]{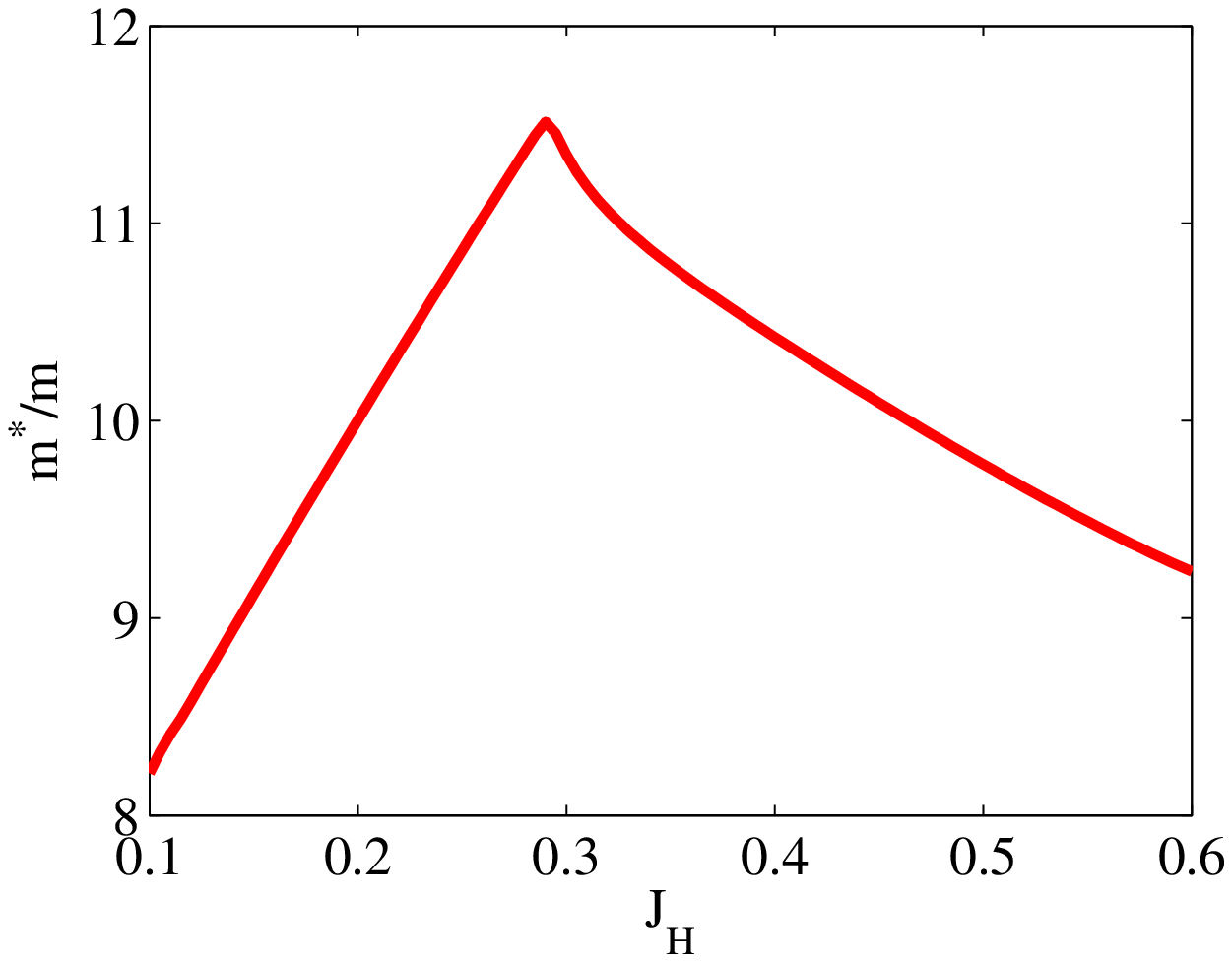}
\includegraphics[width = 0.45\columnwidth]{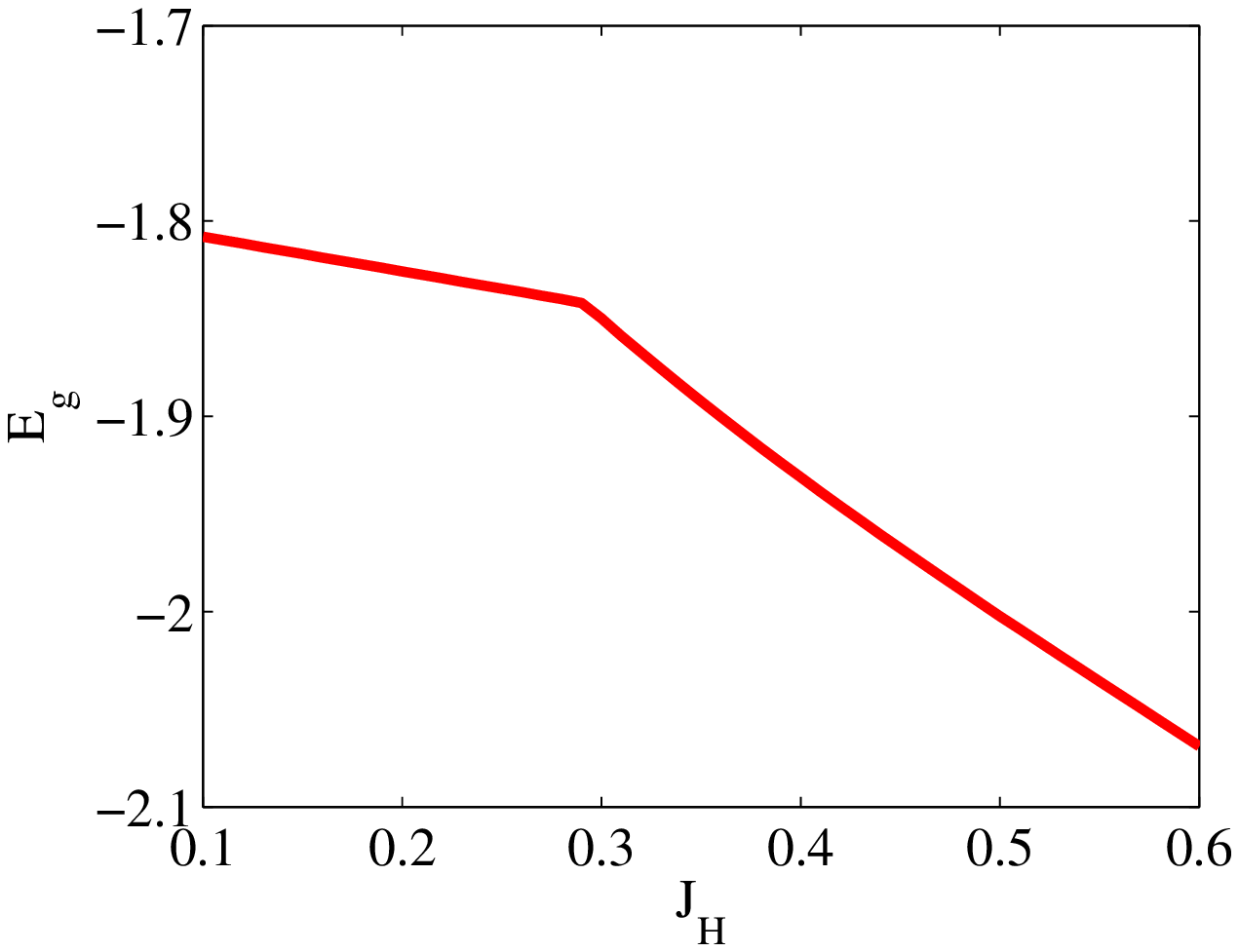}
\caption{\label{fig4} The quasiparticle effective mass $m^{\ast}$ and ground-state energy $E_{g}$ versus $J_{H}$ imply a first-order Lifshitz transition at $J_{H}=0.291$.}
\end{figure}

When $J_{H}=0.2$, a small hole Fermi surface appears around $(0,0)$, which is reminiscent of the bare conduction electron band $\varepsilon_{k}$. Note that the hole Fermi surface is formed when $E_{k}^{-}$ band crosses the Fermi energy. Then, if we further increase the strength of Heisenberg interaction, the quasiparticle $E_{k}^{+}$ band starts to drop into the filled Fermi sea and
the electron-like Fermi surface centered $(\pi,\pi)$ emerges when $J_{H}>0.291$, which signals a radical change of topology of Fermi surface, the Lifshitz transition.\cite{Zhang2011} It should be noted that such transition cannot be described by conventional Landau order parameter. But as seen in Fig. \ref{fig4}, the effective mass $m^{\ast}$ of quasiparticle, which is proportional to the weigh-factor $\beta_{k}^{2}$ averaged over all points on Fermi surface, and the ground-state energy $E_{g}$ provide an explicit signal for such featureless quantum phase transition. Interestingly, the calculated effective mass ranged from 8 to 12 times bare electron mass is well consistent with the quantum oscillation measurement (from 9 to 20 times bare electron mass) in CeCoIn$_{5}$ at its normal state.\cite{Hall2001}
Moreover, the Lifshitz transition here is first-order since the the first-order derivative of the ground-state energy $E_{g}$ at phase transition point $J_{H}=0.291$ is obviously discontinues. As for realistic heavy fermion materials, Lifshitz transition, which could be driven by external applied magnetic or pressure,\cite{Vojta2011} should be manifested in both transport measurement like Hall coefficient and in direct imaging by photoemission spectroscopy.

Therefore, we may summarize that generically a positive valence bond order leads to two Fermi surface structure, which is qualitatively similar to the experimentally observed results in ARPES and QPI.\cite{Koitzsch2009,Jia2011,Yazdani2012,Morr2014} And the corresponding effective mass enhancement agrees with the quantum oscillation experiment. Although the realistic Fermi surface of CeCoIn$_{5}$ is more complicated than our simplified theoretical consideration, the finding here is helpful to give more insight into this exotic heavy fermion superconductor.

\section{Observables in possible superconducting state}\label{sec3}
After obtained qualitatively the correct fermionology of the Kondo-Heisenberg model, it is interesting to check the physical quantities based on the multi-fermi surface topology in the possible superconducting state since the topology of the Fermi surface plays a fundamental role in determining the physical properties of the system. This study is also realistic since CeCoIn$_{5}$ has a superconducting instability below 2.3K. Motivated by the observation that the main contribution comes from the electron Fermi surface centered $(\pi,\pi)$,\cite{Davis2013,Yazdani2013,Morr2014} the extended $s$-wave pairing structure is not favored since it requires more active bands to cancel out the repulsive interaction in intra-band.\cite{Coleman} Therefore, we can safely focus on the other pairing symmetry, i.e. the $d_{x^{2}-y^{2}}$ allowed by the symmetry of square lattice.\cite{Kotliar1988} (A recent dynamical cluster approximation study also finds the clue of $d_{x^{2}-y^{2}}$-wave in the frustrated two-dimensional periodic Anderson model.\cite{Wu2014}) For the present Kondo-Heisenberg model, it has been proposed that the Heisenberg term can induce the pairing between conduction electrons via the pairing of local spins.\cite{Liu2012,Liu2012,Asadzadeh2014} Here, we will follow their formalism and only present basic formula, details on mean-field equations can be found in Ref.\cite{Liu2012}.

\begin{figure}
\includegraphics[width=0.80\columnwidth]{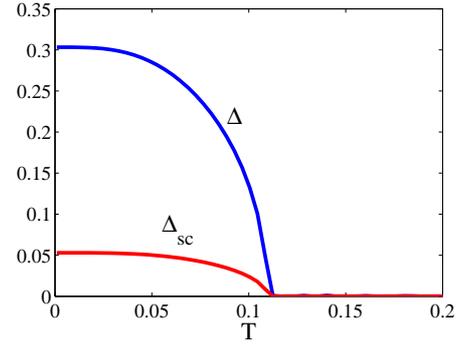}
\caption{\label{fig:8} The pairing strength of local spins ($\Delta$) and conduction electron ($\Delta_{sc}$) versus temperature T.}
\end{figure}

\begin{figure}
\includegraphics[width=0.45\columnwidth]{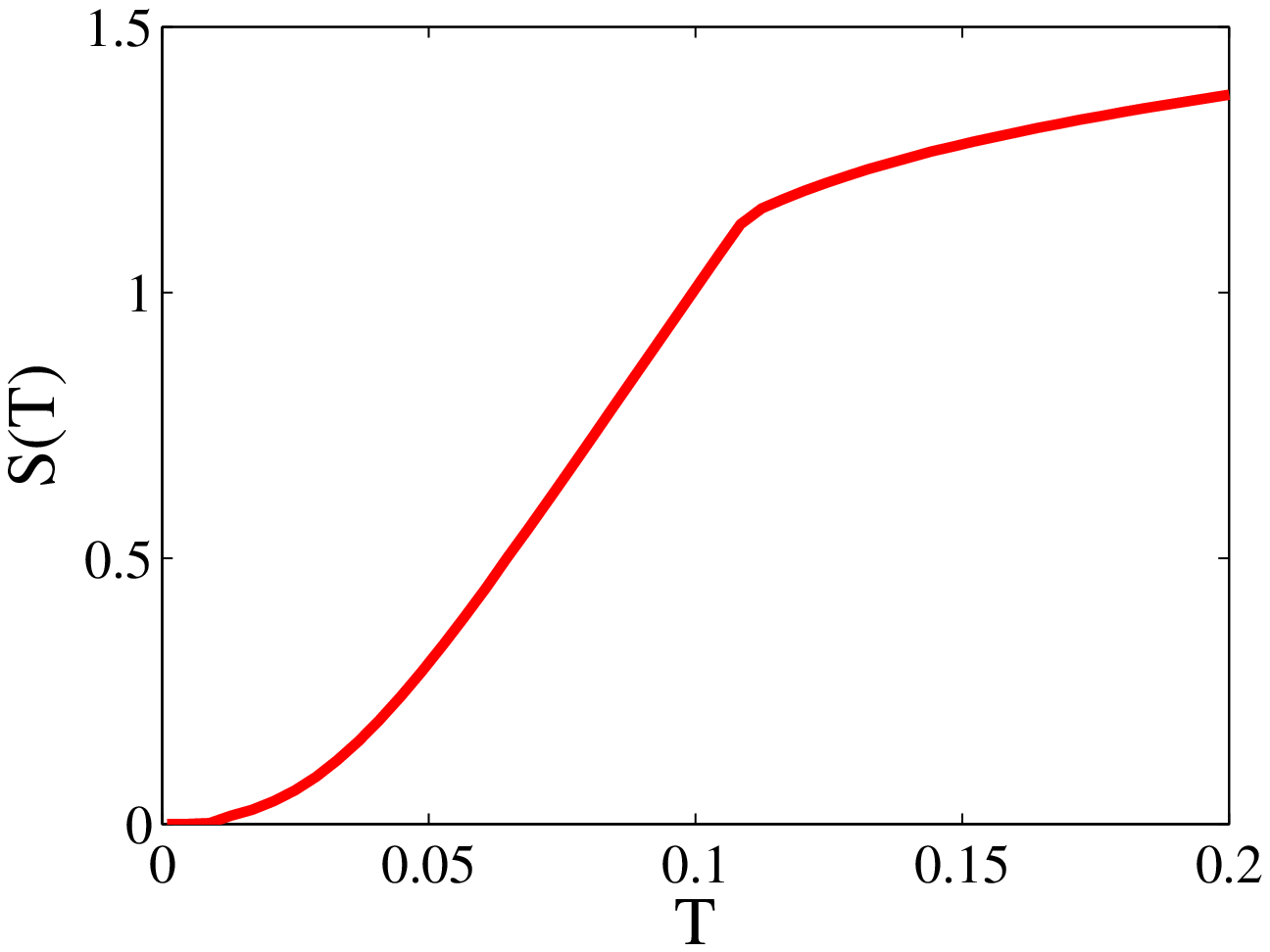}
\includegraphics[width=0.48\columnwidth]{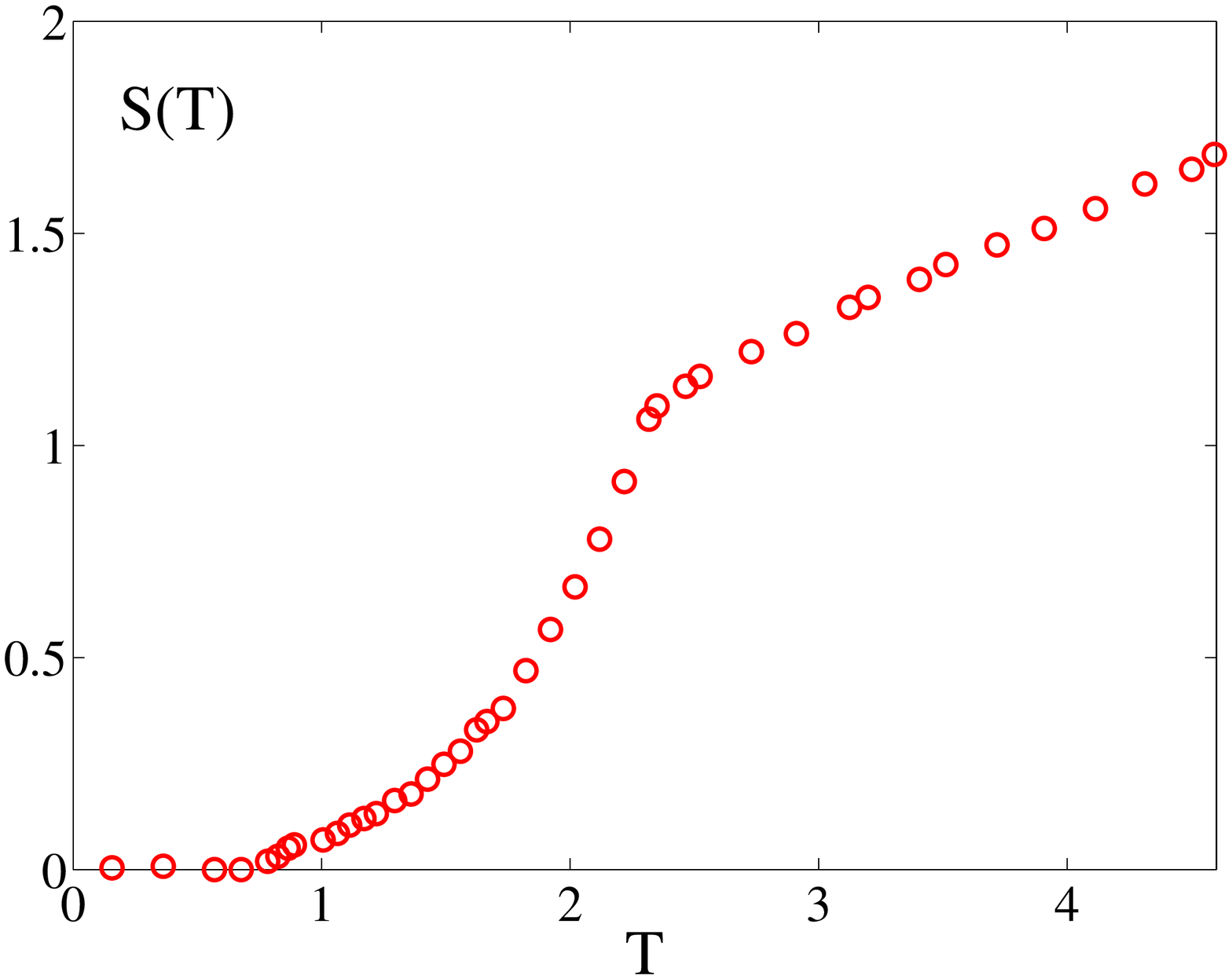}
\caption{\label{fig:9} (Left) The calculated thermodynamic entropy $S(T)$ of the superconducting state versus temperature T. (Right) Entropy
in the superconducting states for CeCoIn$_{5}$ in Ref.\cite{Monthoux2001}. }
\end{figure}

\begin{figure}
\includegraphics[width=0.49\columnwidth]{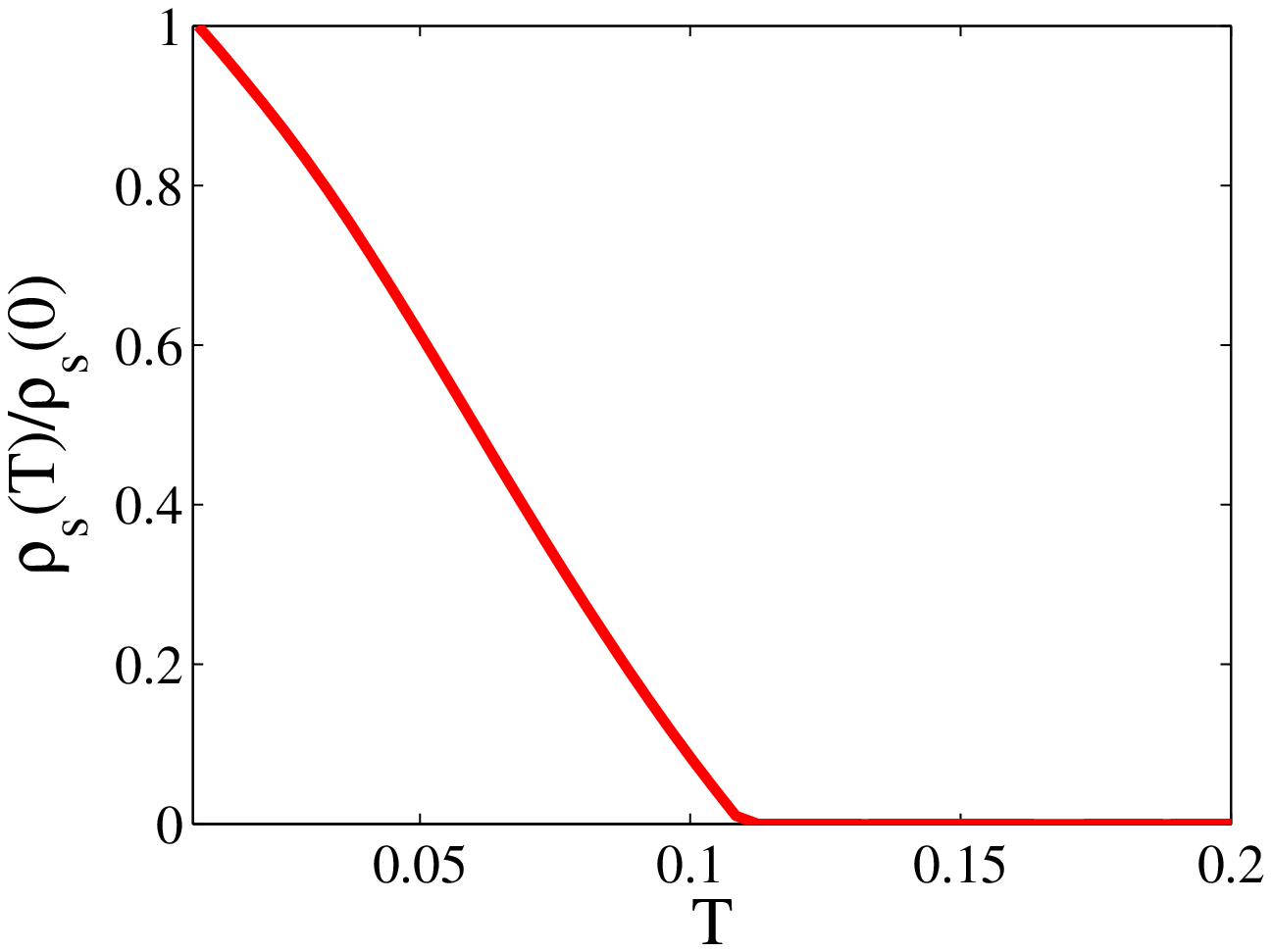}
\includegraphics[width=0.465\columnwidth]{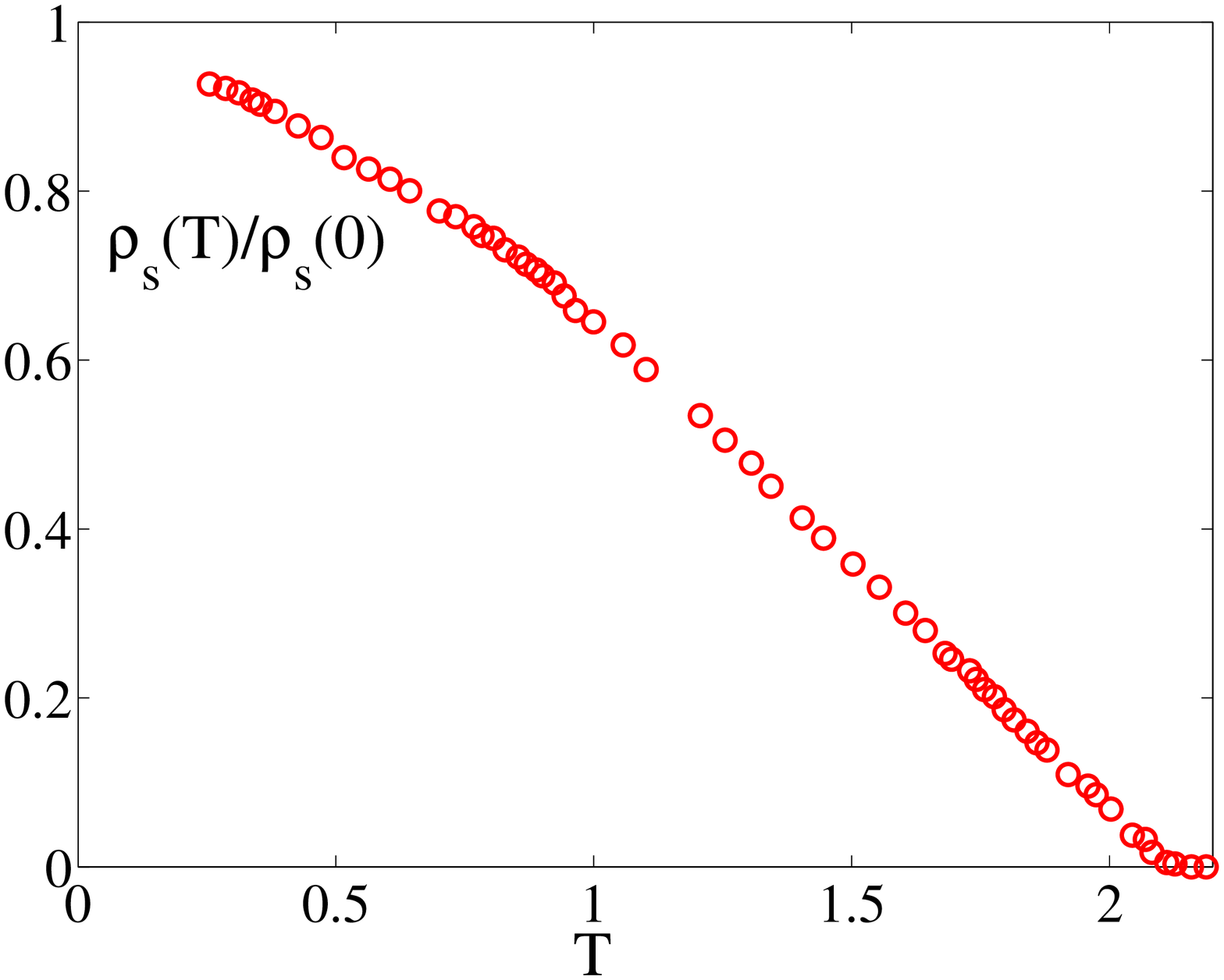}
\caption{\label{fig:10} (Left) Calculated normalized superfluid density $\rho_{s}(T)/\rho_{s}(0)$ in the superconducting state versus $T$.
(Right) The normalized superfluid density $\rho_{s}(T)/\rho_{s}(0)$ of CeCoIn$_{5}$ in Ref.\cite{Ormeno2002}.}
\end{figure}

Like the resonance-valence-bond (RVB) theory for superconductivity in $t-J$-like model,\cite{Zhang1988,Anderson2004} the pairing of local spins contributes a pairing term $J_{H}\sum_{k}\Delta_{k}(f_{k\uparrow}^{\dag}f_{-k\downarrow}^{\dag}+f_{-k\downarrow}f_{k\uparrow})$ with $\Delta_{k}=\Delta \gamma_k$ with $\gamma_k = \cos(k_{x})-\cos(k_{y})$ into the Hamiltonian Eq.\ref{eq1}.
As an example, the evolutions of order parameters, thermodynamic entropy and superfluid density versus temperature are shown in Figs. \ref{fig:8},\ref{fig:9} and \ref{fig:10} with $J_{H}=0.6$.

From Fig. \ref{fig:8}, the pairing strength shows the usual BCS mean-field behavior versus temperature. Although the pairing strength of local spins $\Delta$ is rather large compared to the band width $\sim 4t$, the more realistic pairing strength of conduction electron $\Delta_{sc}$ ($\Delta_{sc}=\langle \sum_{k}\gamma_k c_{-k\downarrow}c_{k\uparrow}\rangle$) has much lower value, which is consistent with the exponential dependence on band width $\Delta_{sc}\sim J_{H}e^{-t/J_{H}}$.
Here, the prefactor $J_{H}$ reflects that the effective pairing is induced by the short-range magnetic fluctuation of Heisenberg exchange interaction.
Besides, we find that the dimensionless quantity, which suggests whether the considered state is a strong coupled superconductor, is $2J_{H}\Delta/T_{c}=5.4$ and $2J_{H}\Delta_{sc}/T_{c}=0.9$. These values imply that for the conduction electron, the system is a weakly coupled superconductor since the pairing is first formed by local electrons and then the Kondo hybridization drives the pairing between conduction electrons. While for local electron, a strongly coupled one is expected due to the performed spin singlet in the RVB background. In heavy fermion superconductor CeCoIn$_{5}$, two pairing strength/gap ($\Delta_{1}\ll\Delta_{2}\simeq 0.6$ meV) have been confirmed,\cite{Pfleiderer2011} and if we consider $\Delta_{sc}$ and $\Delta$ play the role of $\Delta_{1}$ and $\Delta_{2}$, the superconducting instability of CeCoIn$_{5}$ may be driven by the pairing of local electron with "proximity effect" to conduction electron. Similar proximity effect has been suggested as an essential elements in the electron-doped cuparte high temperature superconductor.\cite{Luo2005}

The behavior of thermodynamic entropy and superfluid density is obviously consistent with the standard prediction of $d_{x^{2}-y^{2}}$-wave. Specifically, the quadratic behavior on temperature for low temperature entropy is similar to the finding in original measurement on CeCoIn$_{5}$ as shown in Fig.\ref{fig:9}.\cite{Monthoux2001}
Clearly, such quadratic behavior is due to the gapless nodal quasiparticle from the underlying Fermi surface as studied in last section.
Meanwhile, it can be seen from Fig.\ref{fig:10} that the linear in temperature behavior of superfluid density is also confirmed by the microwave surface impedance measurements in Ref.\cite{Ormeno2002} though more puzzling and controversial power-law behavior exists.\cite{Chia2003,Hashimoto2013} [Such issue may be resolved if the
effective mass effect is isolated from the total
superfluid density as discussed in Sec.\ref{sec5}.]  It is noted that although the system has two superconducting band $E_{k}^{\pm}$, the superfluid density dose not show the intrinsic upward curvature, which is a generic feature of a weakly coupled two-bands system, as first pointed out in Ref.\cite{Xiang1996} for cuprate superconductor. The reason of this difference is that the upward curvature only occurs when the assumed two bands have different critical temperature. However, there is only one critical temperature in our case, thus there is no need to use a weakly coupled two-band picture for our model. For the case of CeCoIn$_{5}$, the intrinsic upward curvature is not observed in London penetration depth measurement,\cite{Ormeno2002,Chia2003} which is correctly covered by our above theoretical calculation and indicates that an effective single-band picture may be useful for understanding some properties in superconducting states.

\begin{figure}
\includegraphics[width=0.80\columnwidth]{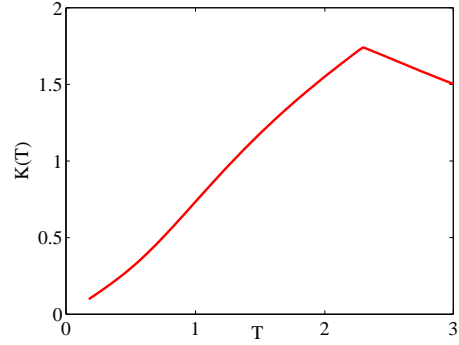}
\caption{\label{fig:11} The Knight shift $K(T)$ versus $T$ in superconducting state.}
\end{figure}

Moreover,the temperature dependent Knight shift $K$ is shown in Fig.\ref{fig:11}, whose linear temperature behaviour is consistent with the prediction of nodal d-wave pairing and experimental measurement in Ref.\cite{Kohori2001}.
\section{STM spectra and quantum interference effect}\label{sec4}
Since recent STM measurement on heavy fermion compounds has provided much invaluable information on the quasiparticle excitation both in normal and superconducting states,\cite{Davis2010,Wirth2011,Yazdani2012,Davis2013,Yazdani2013,Morr2014} in this section, we
proceed to study the STM spectra of the present model.\cite{Yang2009,Coleman2009,Morr2010,Balatsky2010,Vojta2011} Following Ref. \cite{Morr2010}, the zero temperature differential conductance can be obtained as
\begin{eqnarray}
\frac{dI}{dV}=N(\omega)=\frac{2e^{2}}{\hbar}[t_{c}^{2}N_{cc}(\omega)+t_{f}^{2}N_{ff}(\omega)+2t_{c}t_{f}N_{fc}(\omega)]\label{eq6},
\end{eqnarray}
where $N_{cc}(\omega)$,$N_{ff}(\omega)$,$N_{fc}(\omega)$ are local density of state for conduction electron, local electron and the quantum interference term of them, respectively. They are defined by
$N_{cc}(\omega)=\sum_{k}-\frac{1}{\pi}Im G_{cc}(k,\omega)=\sum_{k}[\alpha_{k}^{2}\delta(\omega-E_{k}^{+})+\beta_{k}^{2}\delta(\omega-E_{k}^{-})]$, $N_{ff}(\omega)=\sum_{k}-\frac{1}{\pi}Im G_{ff}(k,\omega)=\sum_{k}[\beta_{k}^{2}\delta(\omega-E_{k}^{+})+\alpha_{k}^{2}\delta(\omega-E_{k}^{-})]$ and $N_{fc}(\omega)=\sum_{k}-\frac{1}{\pi}Im G_{fc}(k,\omega)=\sum_{k}\alpha_{k}\beta_{k}[\delta(\omega-E_{k}^{+})+\delta(\omega-E_{k}^{-})]$. Also, two different tunneling amplitudes $t_{c}$ and $t_{f}$ are introduced for conduction and local electrons.
It has been emphasized in literature that the ratio between these tunneling paths has a strong influence in determining the experimental lineshape of STM spectra,\cite{Morr2010,Balatsky2010} thus we show the corresponding spectra with different ratio of $t_{f}/t_{c}$.

\begin{figure}[!t]
\centering
        \includegraphics[width=0.45\columnwidth]{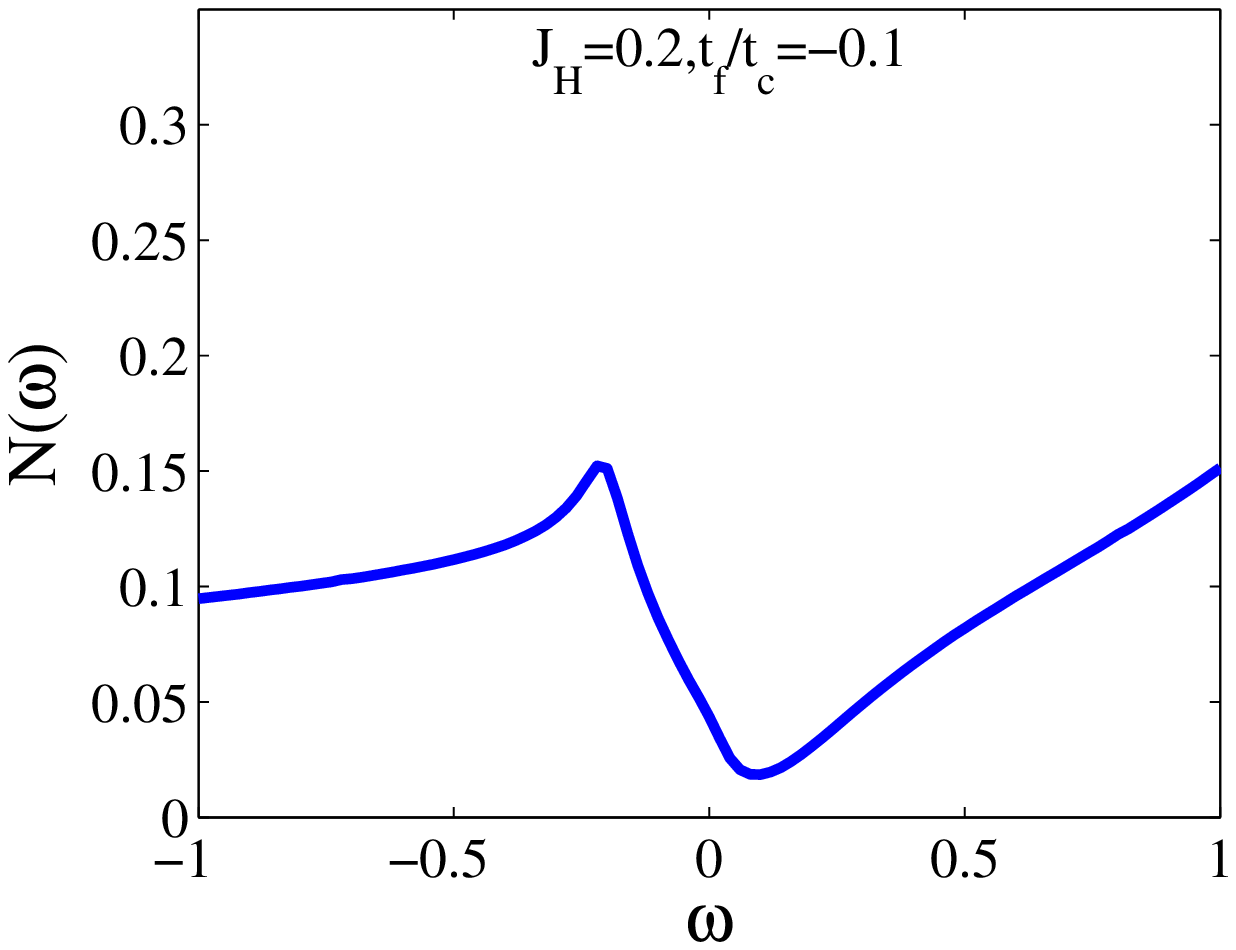}
        \includegraphics[width=0.45\columnwidth]{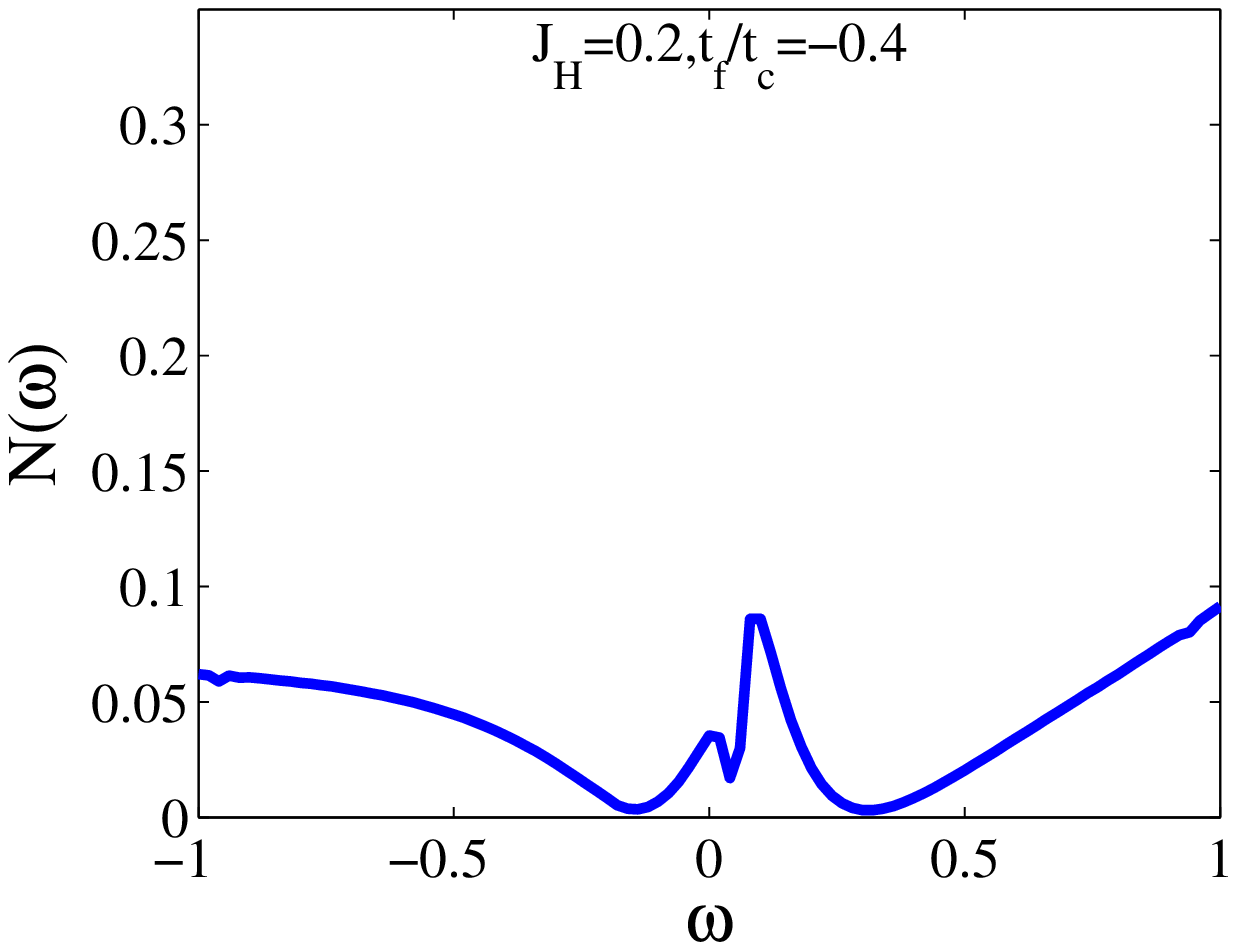}
        \includegraphics[width=0.45\columnwidth]{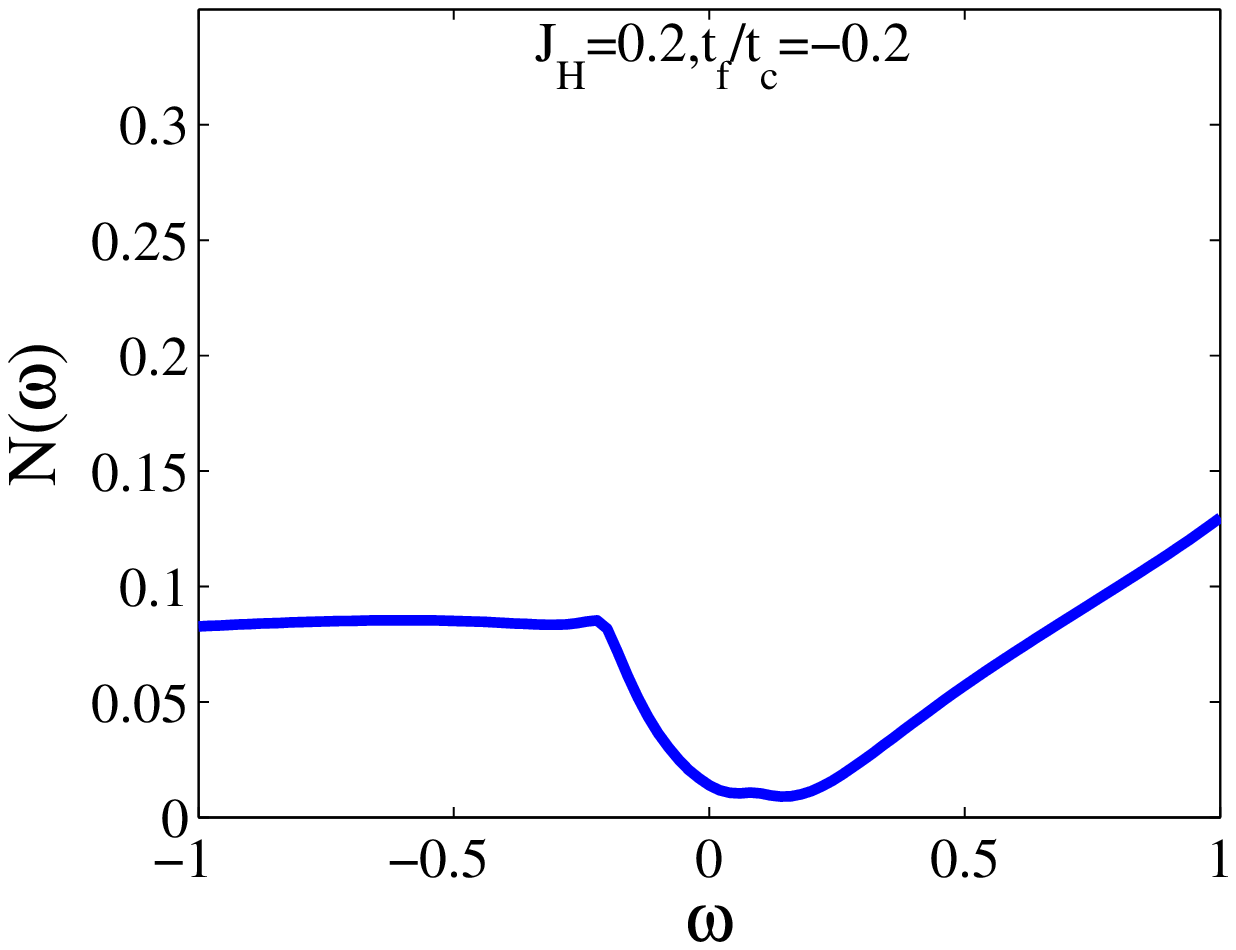}
        \includegraphics[width=0.45\columnwidth]{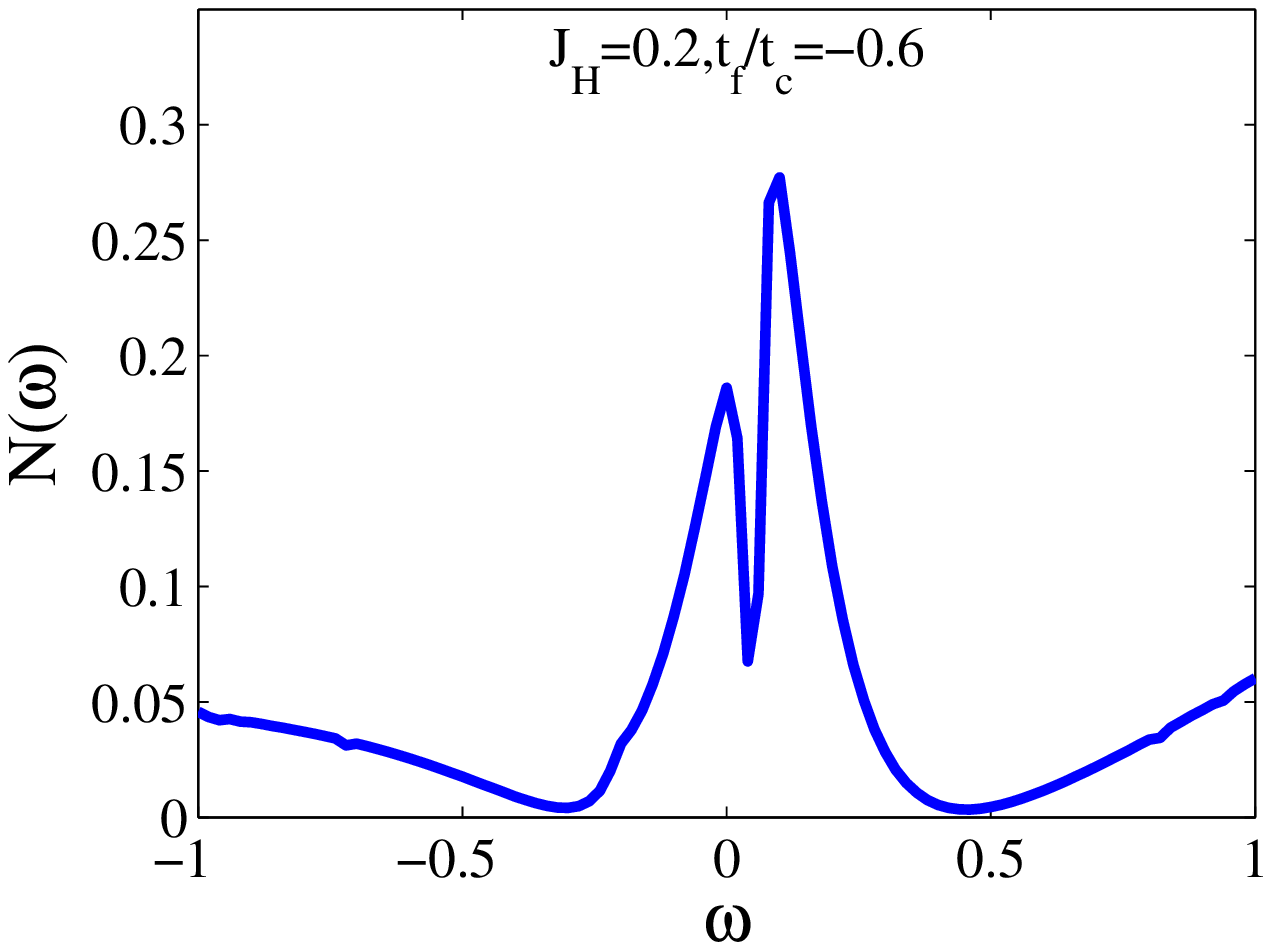}
\caption{\label{fig5}The STM spectrum for single hole Fermi surface with $J_{H}=0.2$.}.
\end{figure}

\begin{figure}[!t]
\centering
        \includegraphics[width=0.45\columnwidth]{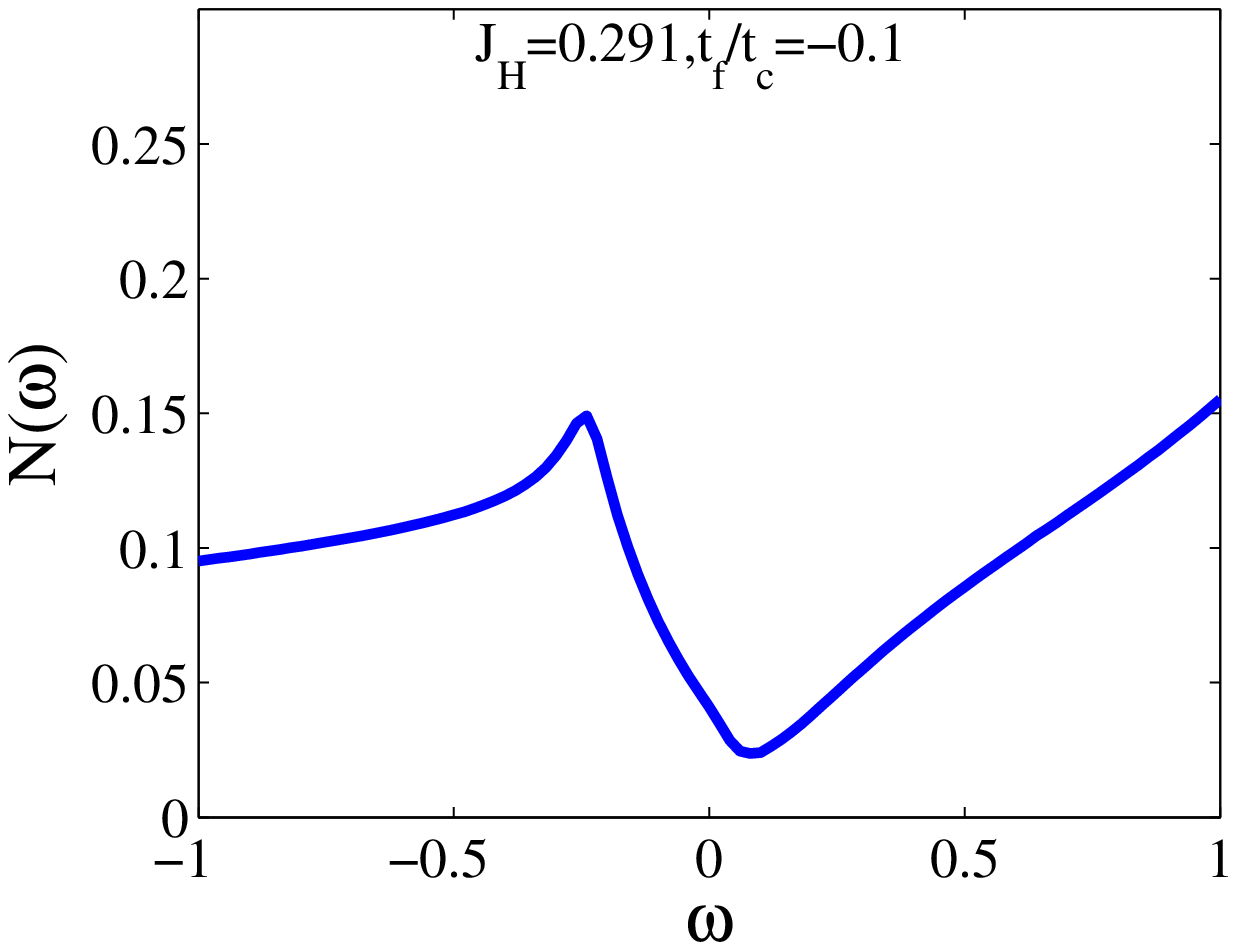}
        \includegraphics[width=0.45\columnwidth]{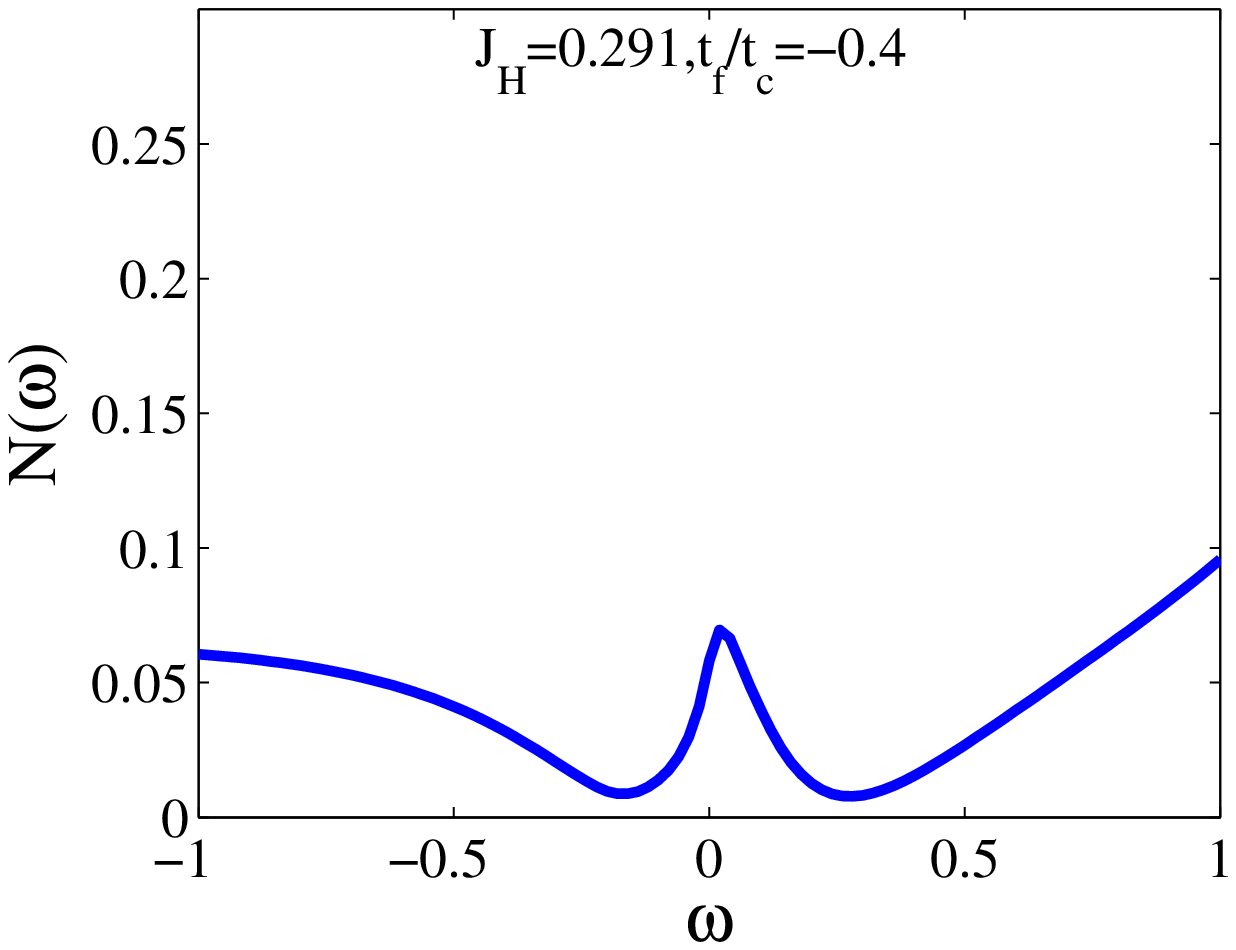}
        \includegraphics[width=0.45\columnwidth]{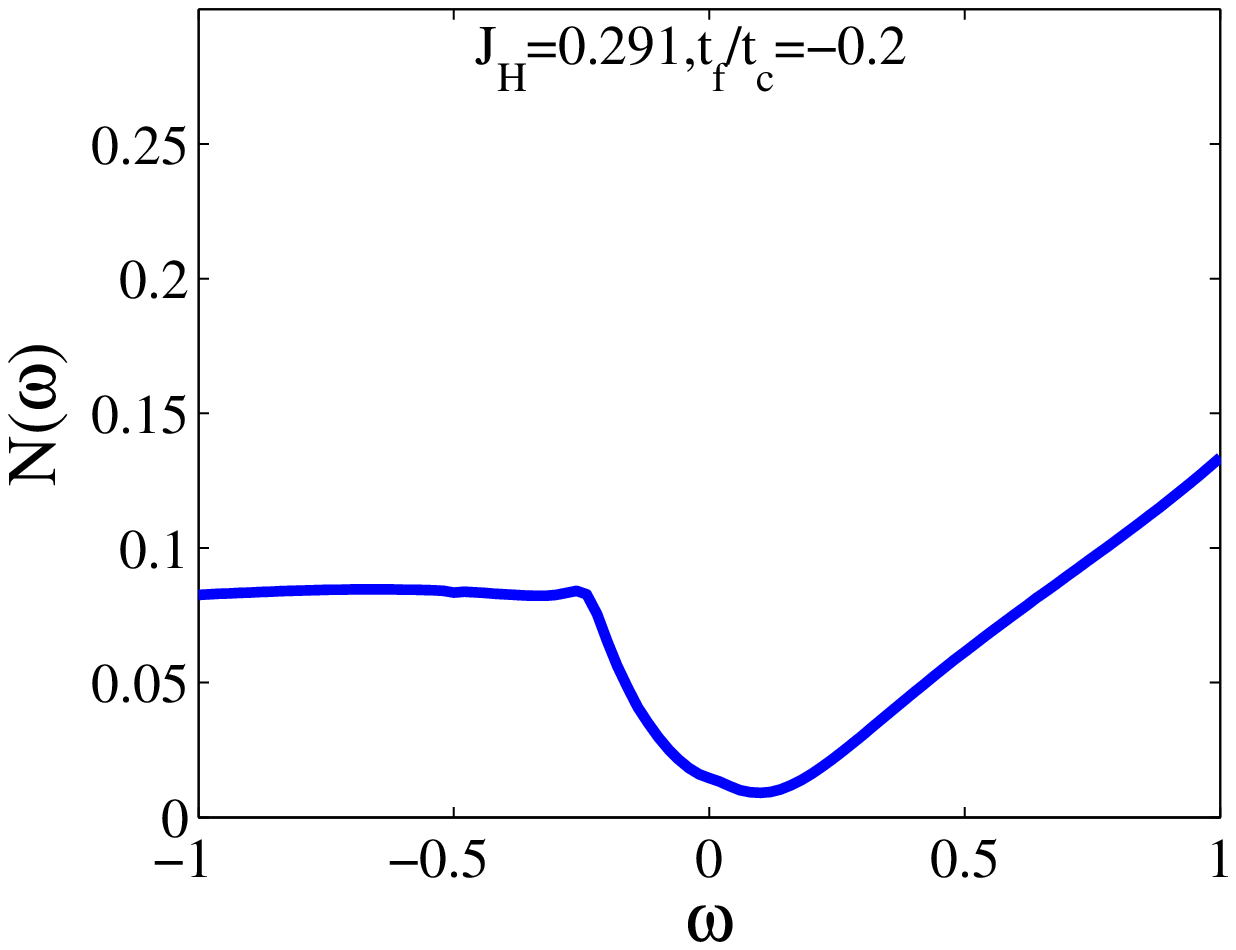}
        \includegraphics[width=0.45\columnwidth]{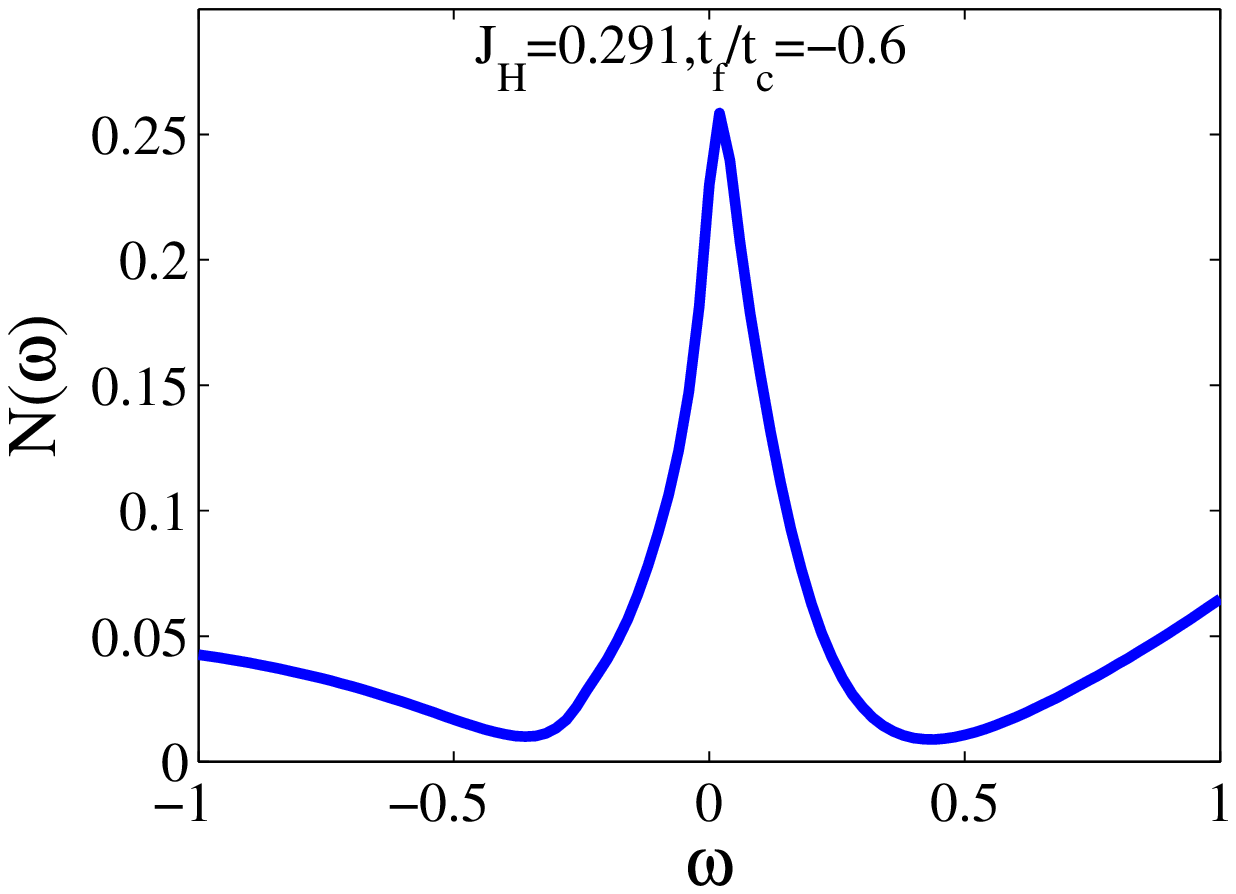}
\caption{\label{fig6}The STM spectrum at Lifshitz transition point $J_{H}=0.291$.}.
\end{figure}

\begin{figure}[!t]
\centering
        \includegraphics[width=0.45\columnwidth]{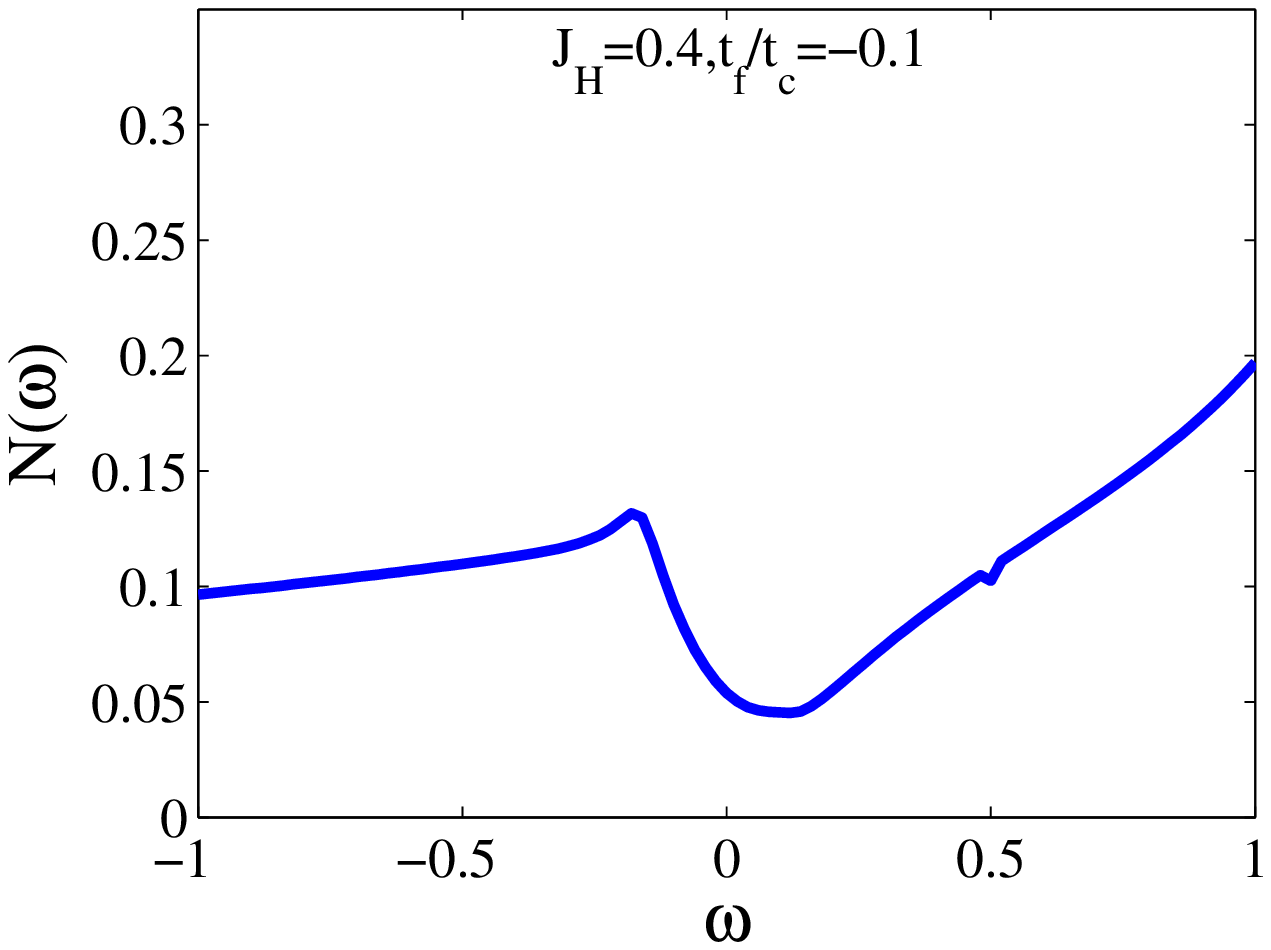}
          \includegraphics[width=0.45\columnwidth]{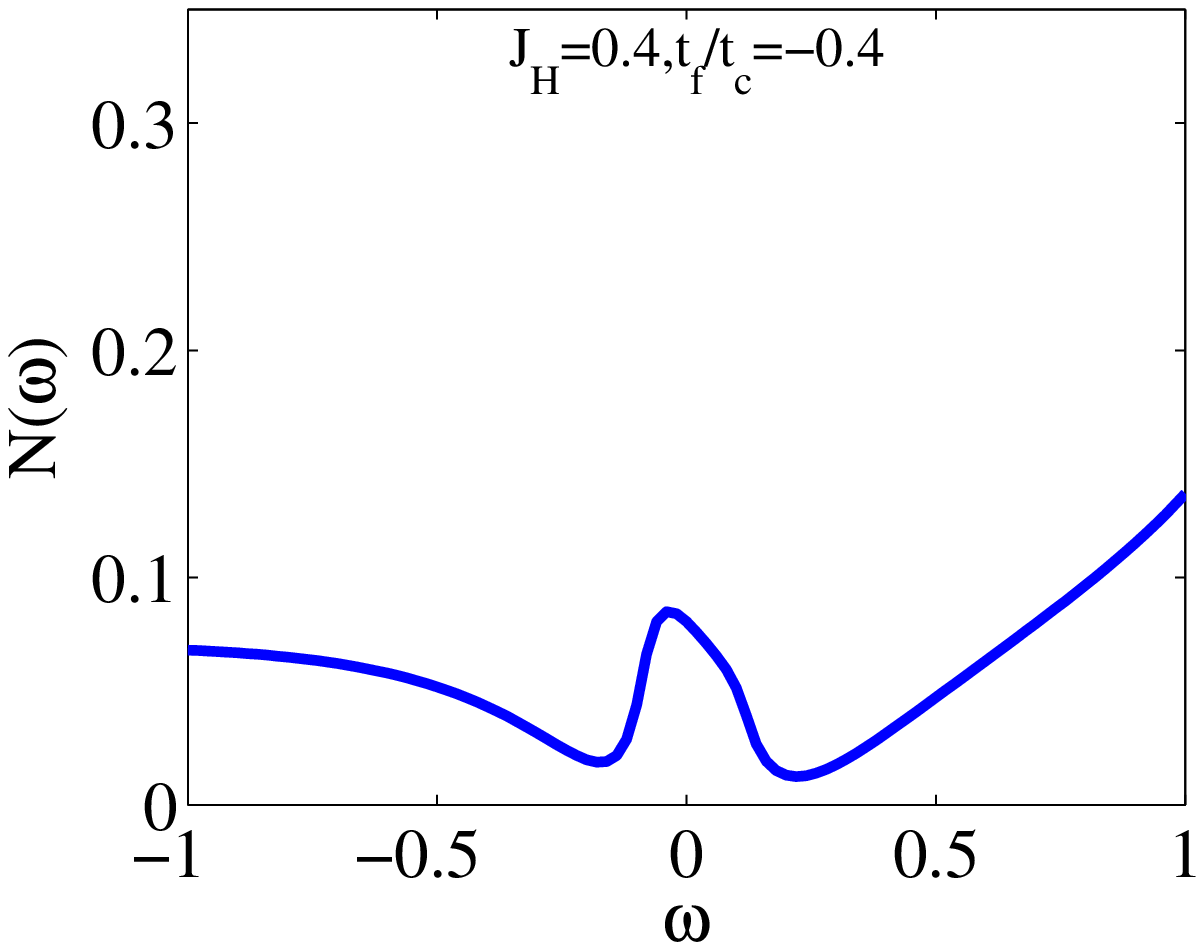}
        \includegraphics[width=0.45\columnwidth]{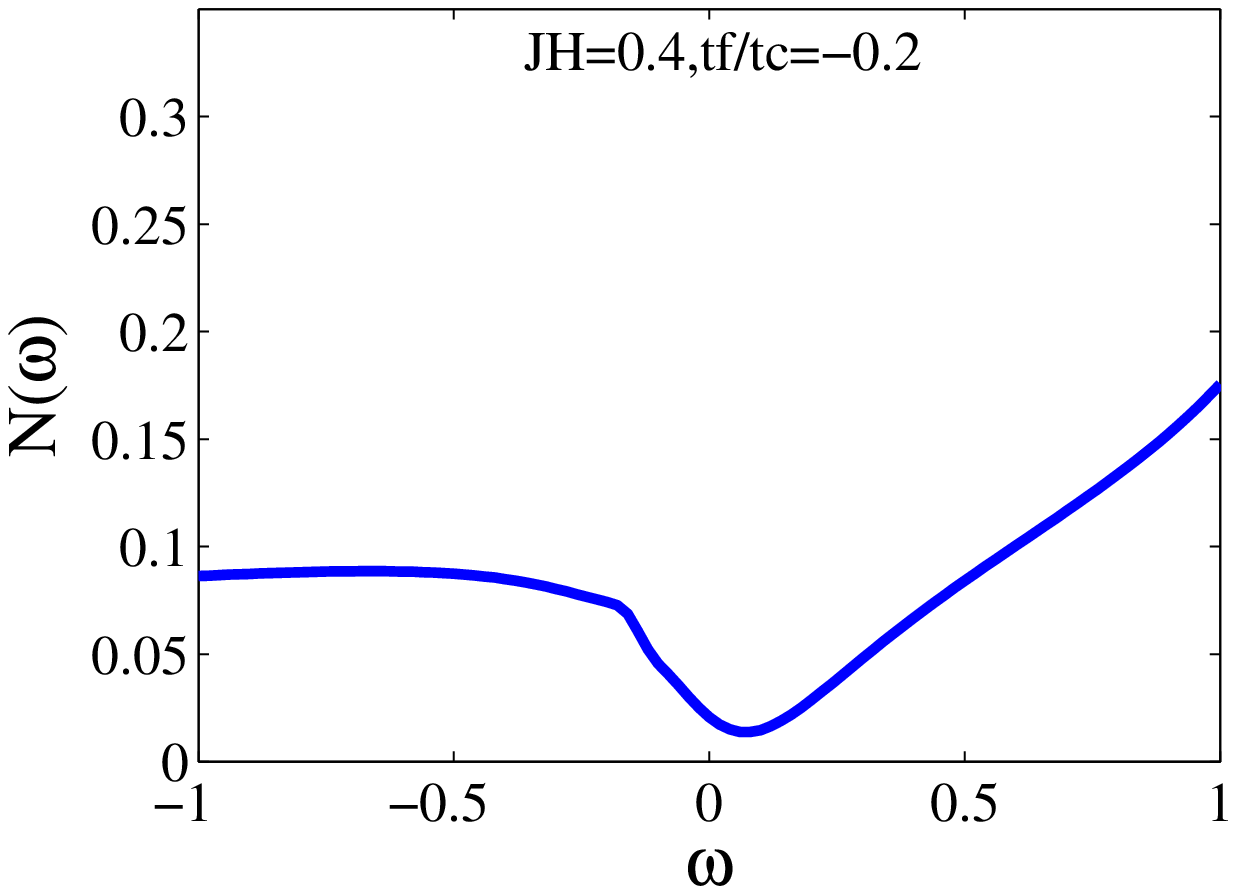}
          \includegraphics[width=0.45\columnwidth]{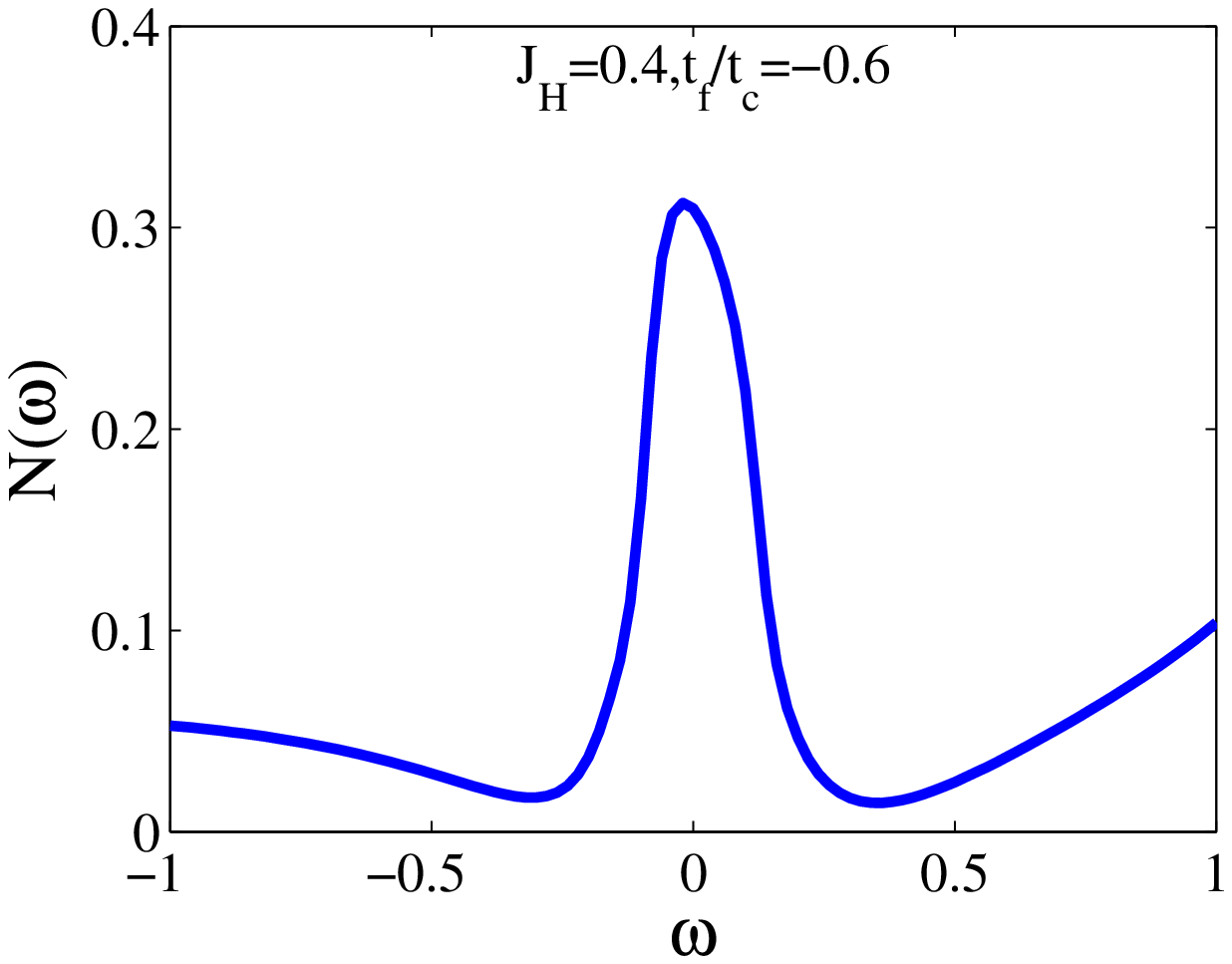}
\caption{\label{fig7}The STM spectrum for two Fermi surface with $J_{H}=0.4$}.
\end{figure}
In Figs. \ref{fig5},\ref{fig6} and \ref{fig7}, we have shown the STM spectra for $J_{H}=0.2,0.291,0.4$. In practice, experimental tunneling results will be modified by the effects of disorder, therefore a phenomenological quasiparticle elastic relaxation rate $\Gamma=0.01$ has been introduced
into the theoretical calculation.\cite{Coleman2009} [More precise treatment can be reached if Fermi liquid theory correction for quasiparticle relaxation rate is considered although the existence of Heisenberg interaction may complicate such issue.\cite{Balatsky2010}] In all cases, it is found that when tunneling is dominated by the conduction electron path ($t_{f}/t_{c}=-0.1$), unambiguous (lattice periodic) Fano lineshape appears which is consistent with the measurement in the so-called hidden order material URu$_{2}$Si$_{2}$ and quasi-two-dimensional heavy fermion superconductor CeCoIn$_{5}$.\cite{Davis2010,Yazdani2012} Next, the classic Fano lineshape is broken by the quantum interference with the local electron path when $t_{f}/t_{c}$ increases. Finally, the
STM spectra are dominated by the tunneling of local electron, which shows a large peak around Fermi energy and is similar to the finding in Ref.\cite{Yazdani2012}. Such peak reflects the fact that there is no excitation gap in this two-band system in contrast to the case with $\chi<0$, where both direct (hybridization) and indirect gap appear in the spectra. It is noted that when the electron Fermi surface emerges, the two-peak or peak-dip-peak structure of single hole Fermi surface evolves into a single peak. Physically, such behavior is caused by filling more states around Fermi energy due to the emergence of the electron Fermi surface. This evolution is consistent with Ref.\cite{Morr2010}, where authors state that the existence of a two peak
structure in $dI/dV$ as predicted in Ref.\cite{Coleman2009} is not a generic feature of heavy-fermion materials.
In our opinion, since only Kondo lattice model is analyzed in Ref.\cite{Coleman2009}, the neglect of possible Heisenberg interaction, which however is crucial for fitting to the observed energy band structure,\cite{Yazdani2012,Morr2014} leads to the flaw in the conclusion of Ref.\cite{Coleman2009}.

\section{Extension and discussion} \label{sec5}
\subsection{On the normal state}
In the main text, we have discussed the Fermi surface structure from the Kondo-Heisenberg model with application to CeCoIn$_{5}$.
However, we should mention that above the superconducting transition temperature, CeCoIn$_{5}$ shows many anomalous behaviors strongly deviated from the prediction of Fermi liquid, e.g. divergent specific heat and linear behavior of resistance over temperature.\cite{Rosch} The model considered in the main text does not include such complexity.

If we want to treat those non-Fermi liquid effect, Kondo-breakdown or spin-fluctuation mechanism should be considered. In those theories, the system is assumed to be close to certain quantum critical point and fluctuation of gapless bosonic modes (from Kondo boson or antiferromagnetic order parameter fluctuation) leads to the non-Fermi liquid-like correction to self-energy and thus results in anomalous behaviors in thermodynamics and transport. \cite{Rosch}
When applied to specific materials like CeCoIn$_{5}$, it is still unknown whether those simplified theories are able to describe the observed complicated phenomena.

\subsection{On the London penetration depth}
In many London penetration depth measurement of CeCoIn$_{5}$, power-law behavior deviated from the linear temperature dependence is observed and seems to violate the explanation of usual d-wave superconductivity.\cite{Chia2003,Hashimoto2013} Interestingly, such issue may be resolved by the idea that the diamagnetic contribution to the superfluid density is indeed temperature-dependent, which reflects that the effective mass is temperature-dependent as close to the quantum critical point.\cite{Truncik2013} When isolating the diamagnetic contribution from the total superfluid density, the paramagnetic contribution is found to have rather well linear temperature-dependence due to nodal quasiparticle.
In our main text, we also find that our model calculation has linear temperature-dependence in low temperature regime. This is because that the underlying normal state is Fermi liquid-like, which automatically kills the anomalous temperature-dependent effective mass and leaves the nodal quasiparticle to be the only active actors.

\section{Conclusion} \label{sec6}
In summary, we have studied the topology of Fermi surface of Kondo-Heisenberg model.
The sign binding is uncovered, which prohibits the formation of multi-Fermi surface structure. When bypassing such difficulty with setting the local electrons dispersion free, we have discovered the evolution of topology of Fermi surface versus the short-ranged antiferromagnetic interaction. Importantly, the obtained Fermi surface is similar to the findings in spectroscopy experiments, thus confirms the validity of our model calculation and arguments.

Furthermore, we have provided the STM spectrum for the discovered multi-Fermi surface system and have studied the physical quantities in the possible unconventional superconducting state. The calculated results are well consistent with existing experiments in heavy fermion superconductor CeCoIn$_{5}$.

In addition, it is suspected that the multi-Fermi surface structure may result from the band-folding effect of preformed antiferromagnetic long-ranged order. However, to our knowledge, the existing
experimental data does not provide explicit information for the antiferromagnetism above the superconducting transition temperature. Both the ARPES and STM results are well explained without any putative magnetic long-ranged order.\cite{Koitzsch2009,Jia2011,Davis2010,Wirth2011,Yazdani2012,Davis2013,Yazdani2013,Morr2014} Therefore, it is still not clear whether the observed multi-Fermi surface is the result of band-folding effect due to certain magnetic order.

We hope the present work may be helpful for understanding on the complicated Fermi surface topology of heavy electron system and the corresponding anomalous behaviors.

\begin{acknowledgement}
Y. Z. and H.-G. L. thank helpful comment from Guang-Ming Zhang and Jian-Hui Dai, who bring us the possible effect of antiferromagnetism to the Fermi surface structure. The work was supported partly by NSFC, PCSIRT (Grant No. IRT1251), the Fundamental Research Funds for the Central Universities and the national program for basic research of China.
\end{acknowledgement}

\section{Appendix}
In the main text, we have discovered the sign binding in the Kondo-Heisenberg model, and it is interesting to see whether such binding exists in other correlated many-body models. Surprisingly, we find that it is indeed true at least in quantum XY and $t-J$-like models in the framework of fermionic large-N method.\cite{Zhang1988,Anderson2004,Yuan2004,Zhong2014} In this Appendix, we take the simplest 1D quantum XY model as an example to check it.
The model is defined by
\begin{eqnarray}
H=t\sum_{i}[\sigma_{i}^{x}\sigma_{i+1}^{x}+\sigma_{i}^{y}\sigma_{i+1}^{y}] \nonumber
\end{eqnarray}
Then, using the same large-N treatment as in the main text,
we obtain the following mean-field Hamiltonian,
\begin{eqnarray}
H=t\sum_{k\sigma}[\epsilon_{k}f_{k\sigma}^{\dag}f_{k\sigma}]+t\chi^{2} \nonumber
\end{eqnarray}
and $\epsilon_{k}=-t\chi\cos(k_{x})$. Next, the only mean-field equation reads
\begin{eqnarray}
\chi=\sum_{k}\cos(k_{x})f_{F}(\epsilon_{k}). \nonumber
\end{eqnarray}

If we only focus on the zero temperature limit, the above equation can be analytically solved, which gives the simple result $\chi=sgn(t)/\pi$. This confirms that the sign binding is true in this simplest 1D quantum XY model.

%
%

\end{document}